\documentclass[12pt,preprint]{aastex}
\usepackage{natbib}
\newcommand{\hr}{\mbox{$^h$}}
\renewcommand{\min}{\mbox{$^m$}}
\renewcommand{\sec}{\mbox{$^s$}}
\renewcommand{\deg}{\mbox{$^{\circ}$}}
\newcommand{\kms}{\mbox{km s$^{-1}$}}

\newcommand{\uJy}{\mbox{$\mu$Jy}}
\newcommand{\ULL}{\mbox{$<27.0$}}
\newcommand{\x}{\mbox{$\times$}}
\newcommand{\Msun}{\mbox{$\rm M_{\odot}$}}
\newcommand{\Mdot}{\mbox{$\dot{m}_{-4}$}}

\newcommand{\E}[1]{\hbox{$10^{ #1 }$}}
\newcommand{\per}[1]{\hbox{#1$^{-1}$}}
\newcommand{\tothe}[1]{\hbox{$^{#1}$}}
\newcommand{\about}{\mbox{$\sim$}}

\newcommand{\nISM}{\mbox{$n_{\rm ISM}$}}

\begin{document}

%%%%%%%%%%%%%%
% Title Page %
%%%%%%%%%%%%%%

\title{{The radio spectra of the compact sources in Arp~220:\\
  A mixed population of supernovae and supernova remnants}}

%\title{THE RADIO SPECTRA OF THE COMPACT SOURCES IN ARP~220:\\
%A MIXED POPULATION OF SUPERNOVAE AND SUPERNOVA REMNANTS}

\author{Rodrigo Parra and John E. Conway} \affil{Onsala Space
  Observatory, SE-439 92 Onsala, Sweden}
% \email{\tt rodrigo, jconway@oso.chalmers.se}
\author{Philip J. Diamond and Hannah Thrall} \affil{Jodrell Bank
  Observatory, Macclesfield, SK11 9DL, UK}
%\email{\tt pdiamond, hthrall@jb.man.ac.uk}
\author{Colin J. Lonsdale} \affil{MIT Haystack Observatory, Westford
  MA 01886, USA}
%\email{\tt cjl@haystack.mit.edu}
\author{Carol J. Lonsdale} \affil{Infrared Processing and Analysis
  Center, California Institute of Technology 100 22 Pasadena CA 91125,
  USA}
% \email{\tt cjl@ipac.caltech.edu}
\and \author{Harding E. Smith} \affil{Center for Astrophysics and
  Space Sciences, University of California, San Diego, La Jolla, CA
  92093-0424, USA}
% \email{\tt hsith@ucsd.edu}

%%%%%%%%%%%%
% Abstract %
%%%%%%%%%%%%

\begin{abstract}
  We report the first detection at multiple radio wavelengths (13, 6
  and 3.6~cm) of the compact sources within both nuclei of the Ultra
  Luminous Infra-Red Galaxy Arp~220. We present the radio spectra of
  the 18 detected sources. In just over half of the sources we find
  that these spectra and other properties are consistent with the
  standard model of powerful Type~IIn supernovae interacting with
  their pre-explosion stellar wind. The rate of appearance of new
  radio sources identified with these supernova events suggests that
  an unusually large fraction of core collapse supernovae in Arp~220
  are highly luminous; possibly implying a radically different stellar
  initial mass function or stellar evolution compared to galactic
  disks. Another possible explanation invokes very short
  (\about3\x\E{5}~year) intense (\about\E{3}~\Msun~\per{year}) star
  formation episodes with a duty cycle of \about 10\%. A second group
  of our detected sources, consisting of the brightest and longest
  monitored sources at 18~cm do not easily fit the radio supernova
  model. These sources show a range of spectral indexes from $-0.2$ to
  $-1.9$. We propose that these are young supernova remnants which
  have just begun interacting with a surrounding ISM with a density
  between \E{4} and \E{5}~cm\tothe{-3}. One of these sources is
  probably resolved at 3.6~cm wavelength with a diameter 0.9~pc. In
  the western nucleus we estimate that the ionized component of the
  ISM gives rise to foreground free-free absorption with opacity at
  18~cm of $<0.6$ along the majority of lines of sight. Other sources
  may be affected by absorption with opacity in the range 1 to 2.
  These values are consistent with previous models as fitted to the
  radio recombination lines and the continuum spectrum.

\end{abstract}


\section{INTRODUCTION}\label{introduction}
Studies of Radio Supernovae (RSNe) and Supernova Remnants (SNRs) in
starburst galaxies can provide unique information about stellar
evolution and Interstellar Medium (ISM) properties in regions
completely hidden by dust at optical/IR wavelengths. Theoretical
models of the radiation emitted by the interaction of the supernova
(SN) blast-wave with the circumstellar medium (CSM) or the local ISM
\citep{CHEVALIER82a, CHEVALIER82b, CHEVALIER94, CHEVALIER01,
  WEILER_REVIEW} combined with radio multi-wavelength/multi-epoch
studies of such RSNe/SNRs \citep{McDONALD01, NEFF04,
  ALBERDI06,BARTEL02} can be used to trace the physical conditions of
the gas swept up as the SN evolves. Additionally from the rate of
newly observed RSNe estimates can be made of the death rate of massive
stars. Adopting a stellar initial mass function (IMF), and a burst
lifetime longer than that of SNe progenitors, this can be converted
into a star formation rate (SFR) and compared to other estimates. This
intercomparison can set constrains on stellar population properties
such as the shape of the stellar IMF.

For over a decade, the ULIRG Arp~220 has been the object of a high
sensitivity VLBI campaign at a wavelength of 18~cm to observe the
evolution of a cluster of unresolved radio sources initially
identified as Type~IIn RSNe \citep{SMITH98,LONSDALE06}. In this paper
we present new observations in which we detect these sources for the
first time at wavelengths shorter than 18~cm (i.e. at 13, 6 and
3.6~cm) at mJy levels. Previous attempts to observe these sources at a
wavelength of 6~cm \citep{ROVILOS05} reported nothing above a
5$\sigma$ level of 250~\uJy. It seems that these earlier observations
failed because a too distant phase calibrator was used (see
Section~\ref{se:astrometry}).  Now that we have made short wavelength
detections we can for the first time study the source spectra and
check for consistency with SN/SNR models, in particular checking
whether the extreme starburst environment of Arp~220 modifies their
properties compared to SNe/SNRs in galactic disks.

The plan of the paper is as follows. In \S 2 we discuss the new
observations and the reduction of the data. \S 3 presents our primary
observational results, including the source spectra, structures and
the spatial distribution of the compact sources. In \S 4 we discuss
the source spectra in more detail and fit simple synchrotron plus
free--free absorption (FFA) models to these spectra.  In \S 5 in light
of new and published information we discuss physical models for the
compact sources, comparing in particular to the expectations of SN and
SNR models. We also discuss in detail the various subclasses of
sources we have identified, set limits on the properties of the
ionized ISM and speculate about their implications for stellar
evolution and the SFR. Finally in \S 6 we summarize our conclusions
and discuss future prospects.

\section{OBSERVATIONS AND DATA REDUCTION}\label{observations}

\subsection{6~cm Global VLBI}
\label{glob6cm}

Two of us (RP and JC) observed Arp~220 at 6~cm on 2005 February 27
with a short snapshot VLBI observation as part of the EVN project
EP049 which is a survey of a large sample of infrared selected
galaxies known as COLA \citep{COLAS1, COLAS2, BADHONNEF}. These
observations involved the Effelsberg (Eb), Westerbork (Wb) and Arecibo
(Ar) telescopes in one of the first user experiments scheduled at a
data rate of 1~Gbit~\per{s}. The full results of this survey will be
given elsewhere (see Parra et al., in preparation). Like all other
sources in the sample, Arp~220 was observed for only 10 minutes.
Despite the short integration time the combination of the high bit
rate and very large telescopes resulted in an untapered noise level of
only $\sigma=38~\uJy$ on the Ar--Eb baseline.

\begin{figure}
\includegraphics[width=1\hsize]{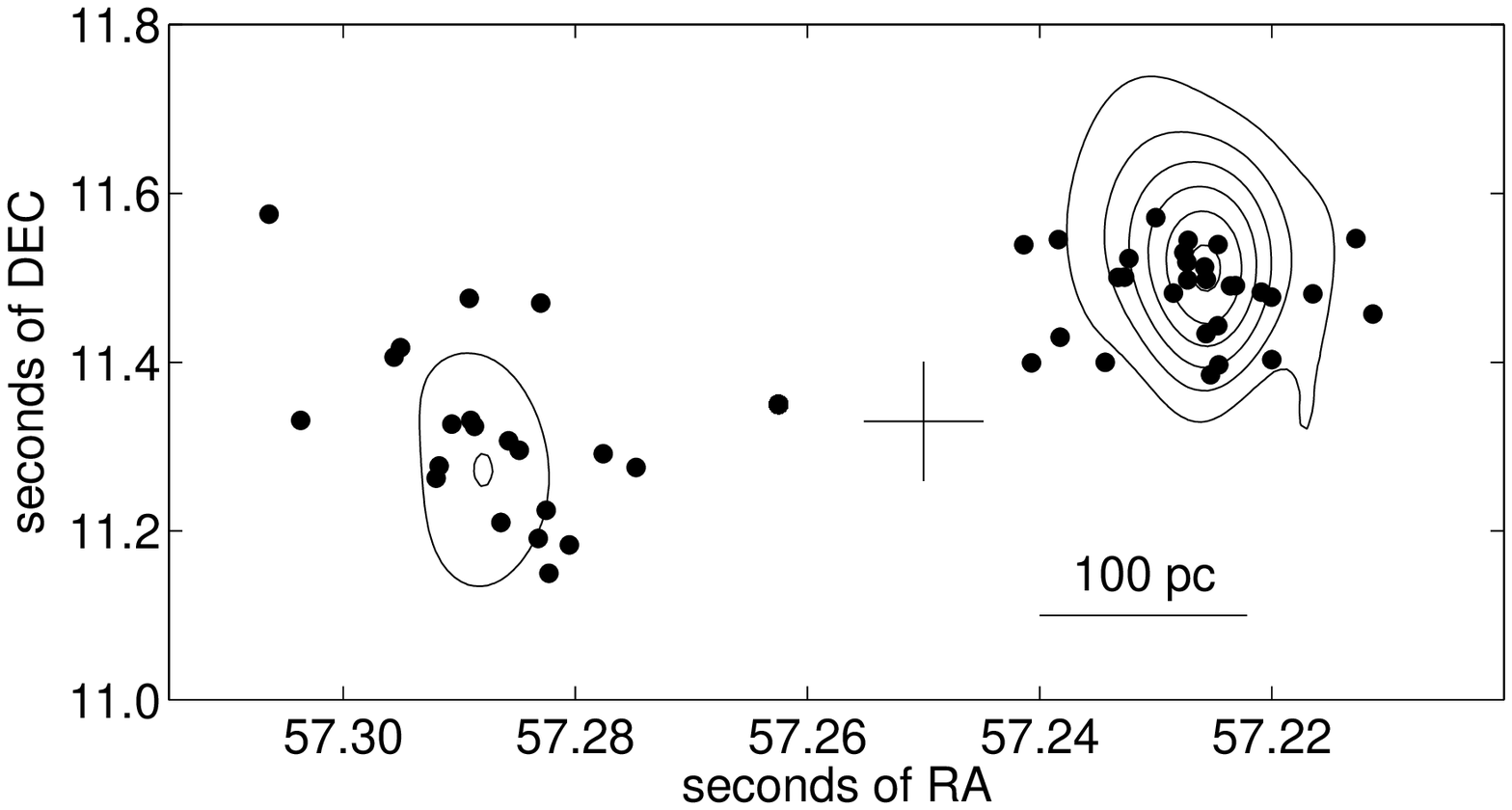}
\caption{Overall view of the Arp~220 nuclear region. Filled circles
  are the corrected positions (see Section~\ref{se:astrometry} for
  details) of the 49 compact sources at 18~cm catalogued by
  \citet{LONSDALE06}. Contours show a 6~cm delay--rate map (``Single
  baseline snapshot image'') obtained from a 10 minute scan with the
  Ar--Eb baseline in project EP049. Contours are drawn 500~\uJy~apart
  starting from 500~\uJy. The data was tapered in time and frequency
  using a Chebyshev window to reduce sidelobes, resulting in a
  delay--rate beam of FWHM \about200~mas and a map noise of
  \about42~\uJy. The cross indicates the reference position
  $\alpha_{2000}=$15\hr34\min57\rlap.\sec25,
  $\delta_{2000}=+$23\deg30\arcmin11\rlap.\arcsec33}
\label{fi:layout}
\end{figure}

The data were correlated on the EVN correlator at JIVE in the
Netherlands. Classic AIPS was used for the calibration stage and
fringe fitting resulting in a detection with a signal to noise ratio
of \about60. We then used our own software to make a delay--rate map
\citep[see][and references therein]{INT_BOOK} from the Ar--Eb data
which showed two sources associated with the eastern and western
nuclei with fluxes of 1.0 and 2.9~mJy respectively (see Figure
\ref{fi:layout}). A delay--rate map made from the Ar--Wb data was
similar but had larger noise. The significance of these maps is that
although they appear to have low resolution they are in fact only
sensitive to structures which are compact on the Ar--Eb
baseline.\footnote{The delay--rate map is a representation of the true
  brightness distribution after high pass filtering and convolving
  with a delay--rate beam. Formally it is the result of the following
  three steps (1) Taking the true image $I(x,y)$ and multiplying it by
  $\exp\left\{-2\pi i (u_{o}x + v_{o}y)\right\}$, where $u_{o}$ and
  $v_{o}$ are the $uv$ coordinates of the center of the patch of the
  $uv$ plane sampled by the time and frequency range of the single
  baseline data. (2) The result is then convolved with a large
  delay--rate beam set by the Fourier transform of the sampled $uv$
  patch. (3) The amplitude of the resulting complex image is taken.}
Therefore although for Figure \ref{fi:layout} the delay--rate beam is
large (\about200~mas) the map includes \emph{only} emission
contributed by features which are less then 2~mas in size and so does
not include any of the extended emission seen in previous 6~cm
\mbox{MERLIN} maps \citep{ROVILOS03}.  {\it This image therefore
  provided the first evidence for the detection of the Arp~220 compact
  features at 6~cm}. Given the large bandwidths now spanned by
1~Gbit~\per{s} observations this delay--rate (or 'single baseline
snapshot imaging') technique is a useful method for detecting and
imaging at moderate resolution compact emission in
starbursts.\footnote{The technique has been subsequently applied to
  several other sources within our COLA sample observations (see Parra
  et al. in preparation).}

\subsection{ VLBA 13, 6 and 3.6~cm Follow-up }\label{se:VLBA}

Initially believing that our global 6~cm observations (see Section~
\ref{glob6cm}) had detected two new bright SNe we applied for and
obtained rapid--response VLBA (NRAO\footnote{The National Radio
  Astronomy Observatory is a facility of the National Science
  Foundation operated under cooperative agreement by Associated
  Universities, Inc.}) observations. These observations used dual
polarization at wavelengths of 13, 6 and 3.6~cm at a total data rate
of 256~Mbit~\per{s}\ and were made on 2006 January 9~th (experiment
BP129). To optimize $uv$ coverage the allocated 10 hours were split
into 8 blocks of $\sim$70 minutes. Each block was equally divided into
23 minutes at each wavelength. The data were reduced using classic
AIPS.  After manual flagging using SPFLG, instrumental delay and rates
were removed by fringe fitting the relatively strong (4~Jy) calibrator
J1613+3412 and the solutions applied to the whole experiment.
Atmospheric effects were removed by using phase referencing with 2
minute scans on the nearby calibrator J1532+2344 (0\rlap.\arcdeg55
away from Arp~220) alternating with 3.5 minute scans on the target at
each wavelength \citep{MEMO24,BEASLEY95}. The observed phases were
very stable at all wavelengths. At 3.6~cm typical differences in phase
between adjacent calibrator scans had rms values of \about4\deg, and
visual estimation of the uncertainties in interpolated phase were at a
comparable level. Given the small separation between the target and
the calibrator the effect of angular variations on phase errors should
be negligible, hence we do not expect any appreciable effect on source
amplitudes due to phase incoherence.  Amplitude calibration was based
on system temperature measurements for each antenna provided as TY
tables; based on previous experience with the VLBA we estimate these
should be accurate to \about5\%.

At each of the three wavelengths the task IMAGR was used to produce
2048\x2048 pixel images of both the western and eastern nuclei regions
with pixel spacing of 0.25~mas~\per{pixel} In order to reduce imaging
artifacts from poorly sampled extended emission a minimum $uv$
distance of 10~M$\lambda$ was used at 6 and 3.6~cm. After
experimentation the most sensitive image at 13~cm was obtained using a
minimum baseline cutoff of 5~M$\lambda$. To maximize image sensitivity
pure natural weighting was used at each wavelength. Only a couple of
hundred CLEAN components were needed before the residual image became
noise-dominated. Despite the use of $uv$ limits the final CLEAN images
contained weak ripples of \about20~mas of wavelength possibly
introduced by the griding algorithm of IMAGR. In a final step these
ripples were removed by taking the Fourier transform of the images,
eliminating the associated spikes and inverse-Fourier transforming
back to obtain almost ripple free images. The final images at 6~cm and
3.6~cm achieved the theoretical noise but the noise at 13~cm was
somewhat larger (see Table~\ref{ta:epochs}). This may be due to to the
remaining effects of extended structure or possibly the weakness of
the calibrator at this wavelength.

Images of the western and eastern nuclei are presented as two RGB
composite images (Figures \ref{fi:1W} and \ref{fi:1E}). These images
were constructed by assigning to the R, G and B channels versions of
the 13, 6 and 3.6~cm images produced by restoring the CLEAN components
using a circular beam of 5~mas at all three wavelengths. These images
clearly show that the detected sources have very different colors
demonstrating their wide variety of spectral properties. In addition
to these images an image was made at 3.6~cm wavelength using uniform
weighting to check for source resolution. All except one of the
sources (see Section~\ref{se:W42}) was unresolved.

\subsection{Additional Data}\label{se:AddData}

In addition to our own BP129 observations we also reduced archival
6~cm VLBA data from 2003 January 2 (BN022) consisting of 2 hours at a
data rate 128~Mbit~\per{s}. These observations used a phase
referencing cycle of 3~minutes on Arp~220 and 1~minute on the
calibrator J1516+1932. These data were reduced following the
procedures described in the previous section.  Finally to aid in the
search for sources at high frequency and for constraining the long
wavelength part of the source spectra we have used data from global
18~cm experiments GD17A \citep{LONSDALE06} and GD17B (Thrall et al.in
preparation). The latter data was observed approximately one year
before BP129, however most 18~cm light curves are known to vary slowly
\citep{ROVILOS05} and so these data can be used to extend the spectra
obtained from BP129. A summary of all the available data indicating
the achieved sensitivities and resolutions is given in
Table~\ref{ta:epochs}.

\subsection{Source Astrometry and Previous 6~cm
  Observations}\label{se:astrometry}

Comparing our 13~cm image from experiment BP129 with published 18~cm
images \citep{SMITH98} we found 4 bright sources with the same
relative separations, however the absolute positions of the two images
differed by \about100~mas in declination. Careful checking convinced
us that the positions from BP129 are correct and that it is the
published positions that are wrong. In particular 6~cm source
positions from EP049, BP129 and BN022 all agree.

Previously published 18~cm VLBI continuum and spectral line images
have all been made by phase-referencing to the brightest 1667~MHz OH
maser in the western nucleus \citep{SMITH98,ROVILOS03}. Absolute
positions were then assigned by adding the absolute position of this
maser as determined from an old MERLIN astrometric observation. The
fact that we find an error in absolute position which is entirely in
one coordinate and is very close to 100~mas suggests that a
typographical error was made when either originally recording or later
manually copying the position of this brightest maser. Note that this
astrometric error effects both the published absolute positions of
18~cm continuum sources and OH masers. It will not of course effect
the relative positions of masers and compact 18~cm continuum.

A consequence of the above astrometric error is that previous searches
for sources at 6~cm were made centered on the wrong field. {\it
  However this does not appear to be the primary reason why previous
  observations at 6~cm failed to detect anything}. The 6~cm data from
\citet{ROVILOS05} has been re-reduced and the correct positions
searched but only hints of detections have been found. The reason is
most likely the large angular distance (5\rlap.\arcdeg76) to the
calibrator combined with the fact that these observations were made
near solar maximum when ionospheric effects could have been
significant even at 6~cm. In contrast, the new absolute positions
derived from our BP129 VLBA observations are accurate to within a few
mas \citep{PRADEL05}. It should be noted that the nearby calibrator
used for these observations was not catalogued at the time of the
earlier imaging attempts at 6~cm \citep{PETROV05}.

\section{RESULTS AND ANALYSIS}\label{se:analysis}
\subsection{Source Detection Criteria}\label{se:detect}

\begin{figure*}
\centering
\includegraphics[width=0.7\hsize]{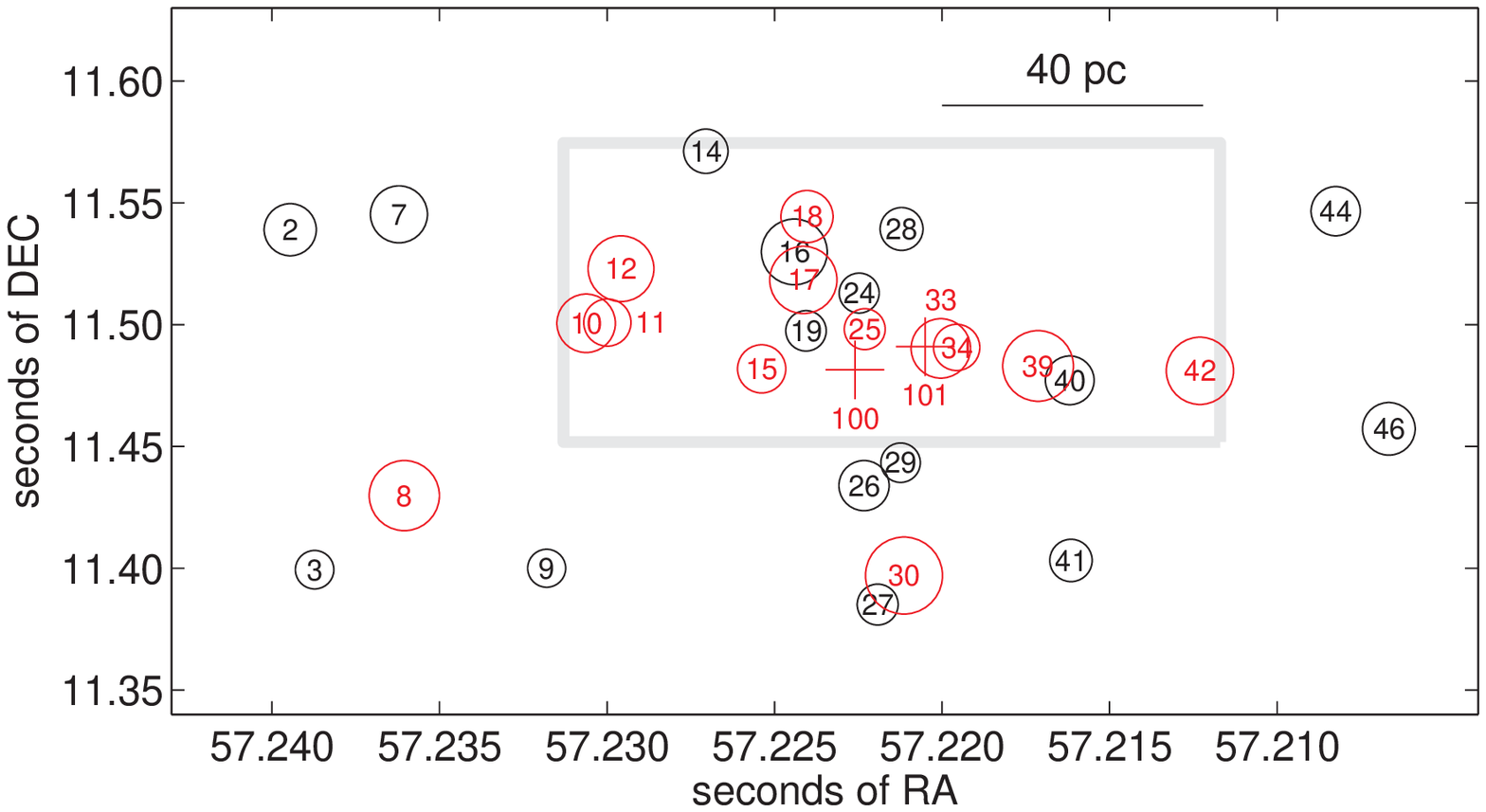}
\includegraphics[width=0.95\hsize]{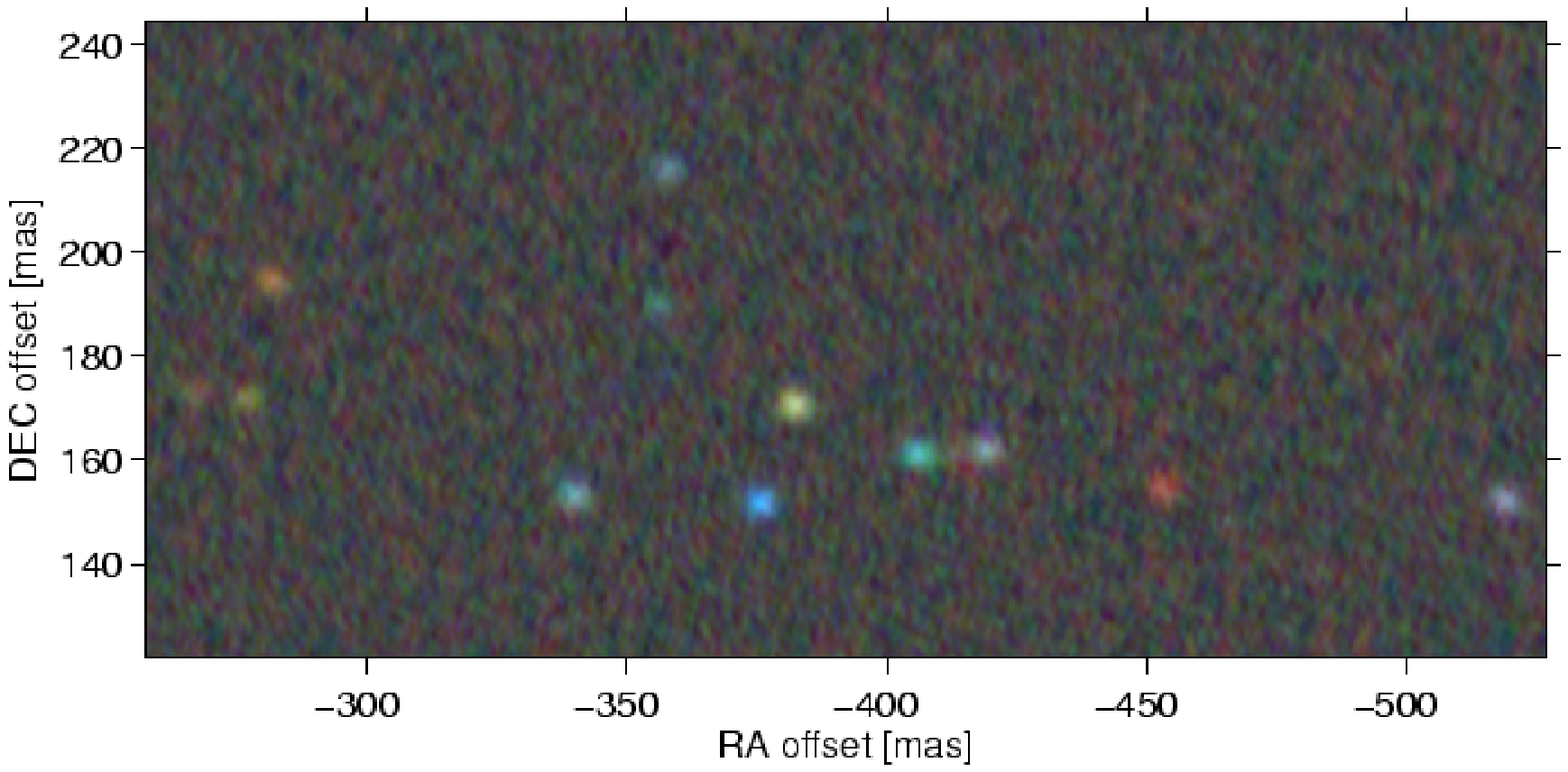}
\caption{The distribution of compact radio sources in the western
  nucleus. {\bf Top}: Detected sources at 18~cm wavelength shown as
  circles with area proportional to flux density (as reference, source
  number 30 is 1.23~mJy). The plotted numbers, prefixed by ``W'' for
  ``West'', are the source names as defined by \citet{LONSDALE06},
  including two new sources (W55 and W56). Sources detected at any
  wavelength shortward of 18~cm are shown in red. Red crosses indicate
  the positions of two previously uncatalogued sources. The gray
  rectangle indicates the relative position of the composite image
  shown in the bottom panel. {\bf Bottom: } Red, Green and Blue
  composite image displaying respectively 13, 6 and 3.6~cm images from
  experiment BP129. The axes of this image are in mas from the
  reference position indicated by the cross in Figure~\ref{fi:layout}.
  An expanded contour image of the region around W33 is shown in
  Figure~\ref{fi:contours} }\label{fi:1W}
\end{figure*}

\begin{figure*}
  \centering
\includegraphics[width=0.45\hsize]{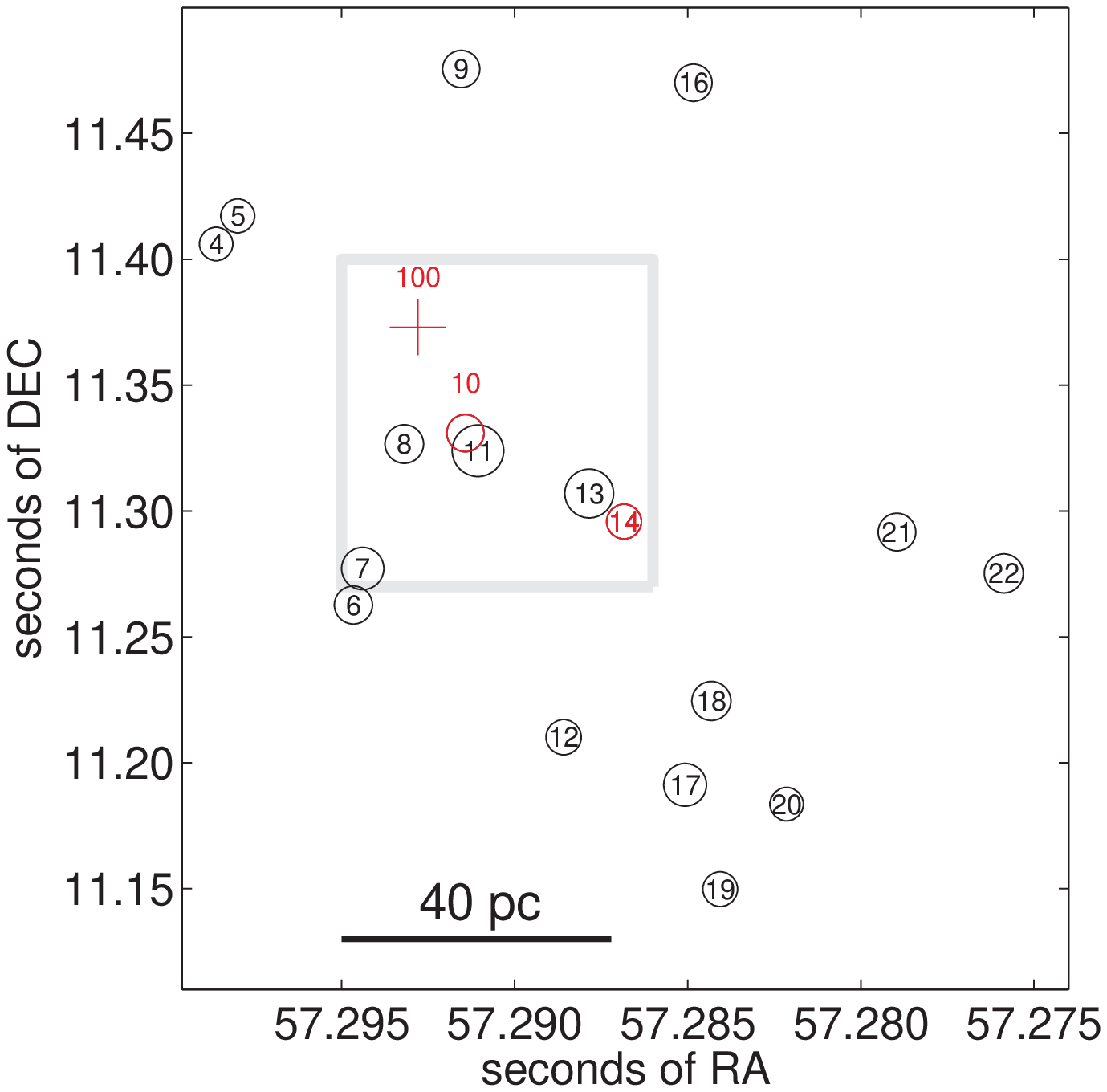}\hfil
\includegraphics[width=0.45\hsize]{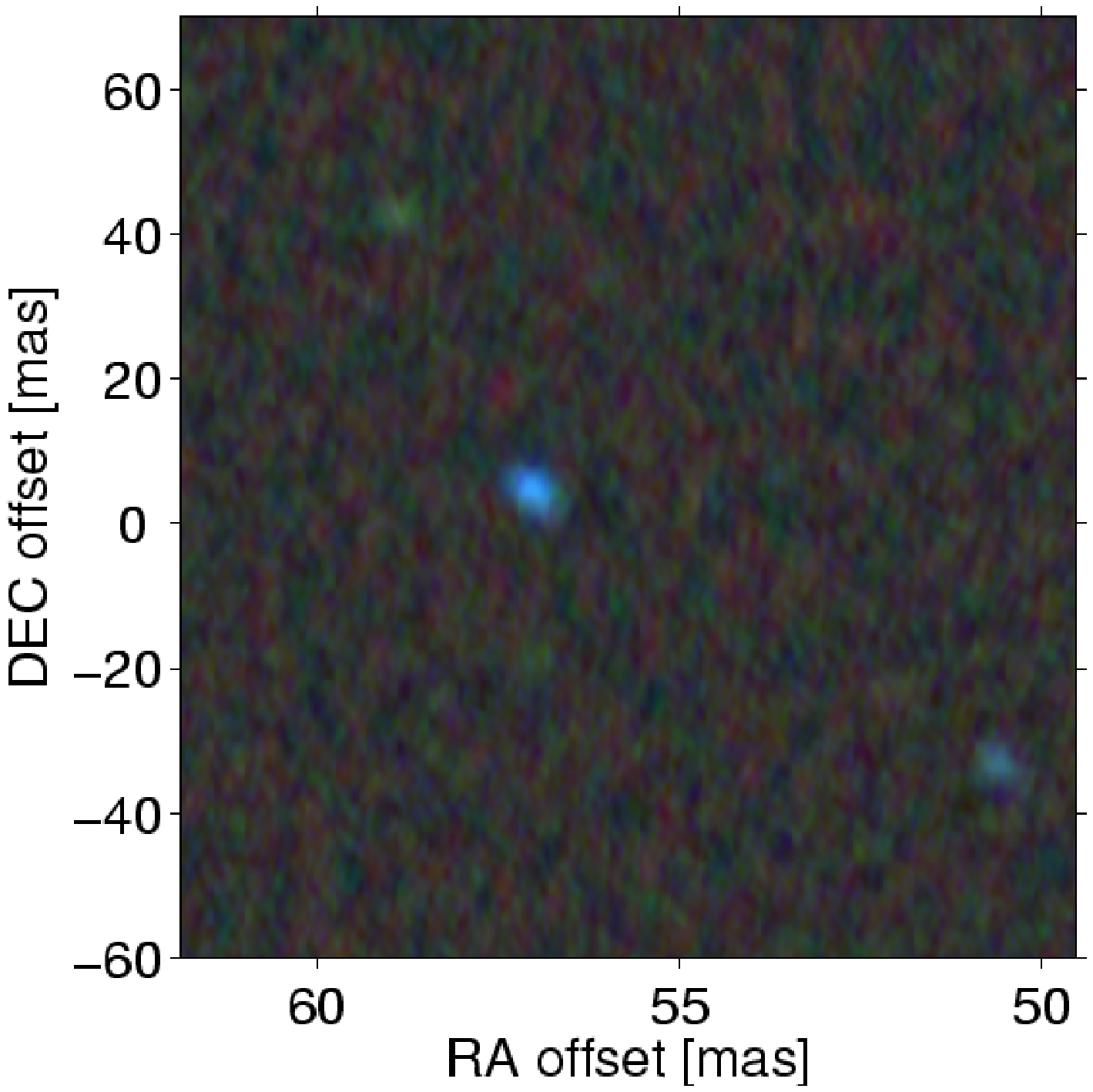}
\caption{The distribution of compact radio sources in the eastern
  nucleus. {\bf Left}: Detected sources at 18~cm wavelength shown as
  circles with area proportional to flux density. The plotted numbers,
  prefixed by ``E'' for ``East'' are the source names as defined by
  \citet{LONSDALE06}, including one new source (E24). Sources detected
  at any wavelength shortward of 18~cm are shown in red. Red crosses
  indicate the position of the previously uncatalogued source.  The
  gray rectangle indicates the relative position of the composite
  image shown in the right panel. {\bf Right: } Red, Green and Blue
  composite image displaying respectively 13, 6 and 3.6~cm images
  BP129. The axes of this image are in mas from the reference position
  indicated by the cross in Figure~\ref{fi:layout}.}\label{fi:1E}
\end{figure*}

Visual inspection of Figures~\ref{fi:1W} and \ref{fi:1E} shows that
many sources are clearly detected at short wavelengths. However to
more rigorously define a complete list of detections we defined
quantitative detection criteria. We made two passes through the BP129
images, first searching around the positions of catalogued 18~cm
sources and then searching the whole area of the eastern and western
nuclei. In the first pass sources could be accepted as real at a lower
signal to noise ratio than in the second pass because of the much
smaller number of beam areas searched.  In the first pass we searched
separately at each of our three wavelengths; a source was considered a
detection if it was detected at one or more wavelengths. Since sources
missed in the first pass would be uncatalogued at 18~cm (and hence
very weak at both 18 and 13~cm) these sources would have spectra that
peak at high frequency. Therefore in the second pass we only searched
a single composite image formed by averaging our 6 and 3.6~cm images.

In the first pass we conservatively took the 18~cm beam size as our
search area around each of the 49 known positions. The number of
independent beam areas searched at each wavelength ($\lambda$) was
then $N_s=49(18/\lambda)^{2}$. Analysis of the histogram of the images
in regions away from detected sources showed distributions that were
very close to Gaussian allowing the use of Gaussian statistics.
Setting the detection criterion as any spike within a given search
area above $\eta\sigma$ the probability that one or more entries in
the list of detected sources is a noise spike is
$F=1-P(I<\eta\sigma)^{N_s}$, where $P(I<\eta\sigma)$ is the cumulative
probability of a noise spike at a given beam area being less than
$\eta\sigma$. At each wavelength we chose $\eta$ such that $F$ was
0.5\%. The resulting detection limits at 13, 6 and 3.6~cm are 3.8, 4.1
and 4.4$\sigma$ respectively. Using these criteria a total of 15
sources were detected in the first pass.

In our second pass we then looked for new sources at any other
position contained within two square fields of 512\x512~mas enclosing
the eastern and western nuclei (see Section~\ref{se:VLBA}). As
explained above we searched a composite image made by averaging the
6~cm and 3.6~cm images. This time the area searched was much larger
than in the first pass, therefore for a 0.5\% confidence of false
detection the required detection limit was $5.1\sigma$. In this second
pass we detected 3 new sources (2 in the west). In total, despite our
rather modest sensitivity, we detected a total of 18 sources (see
Table~\ref{ta:source_fluxes}). Given that in the first pass three
separate searches were made (one at each wavelength) and one search
done in the second pass, all with 0.5\% confidence of false detection,
we estimate an overall 2\% probability that one or more of the sources
in Table~\ref{ta:source_fluxes} is a false detection.
 
It should be mentioned that the BP129 image at 3.6~cm shows several
point like features that lie just below our detection threshold and
are mostly located in the region with the highest density of sources
(within 20~mas of position ($-$380, 160)~mas in the bottom panel of
Figure~\ref{fi:1W}). Almost certainly some of these features will be
confirmed as detections in higher sensitivity observations.

\subsection{Source Structure}\label{se:structure}

The only resolved source at 3.6~cm is W42. This source is located
\about100~mas to the west of the region with the highest surface
density of sources (see Figure~\ref{fi:1W}).  Along the position angle
corresponding to the beam minor axis the FWHM of W42 is about twice as
big as the beam FWHM (see Figure~\ref{fi:contours}-Left). The fact
that other sources were unresolved shows that the resolution of W42 is
not an artifact of residual phase errors in the data.  Another source
deserving special attention is W33 which in addition to its peculiar
spectrum (see Figure~\ref{fi:spectra2}) shows possible signs of an
elongated structure in the North--South direction at 13~cm (see
Figure~\ref{fi:contours}-Right).  This source is tightly located
between two bright sources (W56 and W34) and lies near the edge of a
patch of diffuse 18~cm emission covering the central region of the
western nucleus (Thrall et al. in preparation).  We speculate about
the nature of this source in Section~\ref{se:W33}

\subsection{Source Variability}\label{se:variability}
In order to search for correlations between source spectral properties
and source age we classify our detected sources in terms of their
broad 18~cm variability properties. A more detailed variability study
including detailed light curve modelling will be presented in a
subsequent paper (Thrall et al. in preparation). We define four
classes of sources:

\vskip 0.2cm
\noindent \emph{Class L (Long lived)} These are 8 sources already
detected by \citet{SMITH98} based on observations in 1994, hence these
sources are at least 11~years old. In addition analysis of the first 6
years of monitoring data by \citet{ROVILOS05} showed that all these
sources had slow or undetectable decays of their 18~cm flux densities
and are therefore likely to be significantly older.

\vskip 0.2cm
\noindent \emph{Class R (Rising)} These are 4 sources not detected by
\citet{ROVILOS05} but which are detected in the latest 18~cm epochs at
flux levels significantly above the \citet{ROVILOS05} detection
limits. W11, W25 and W34 were present in GD17A but not in the 18~cm
experiment (GL26B) made 12 months earlier \citep{LONSDALE06}. Source
W11 was present in both GD17A and GL26B but not at earlier epochs
\citep{LONSDALE06}.

\vskip 0.2cm
\noindent \emph{Class S (Single epoch)} These are 3 sources only
detected in our new high frequency data and not in any 18~cm epoch.

\vskip 0.2cm
\noindent \emph{Class A (Ambiguous)} These are sources with ambiguous
variability classification. For instance one of the three sources in
this class (W15) is detected in both GD17A and GD17B at similar flux
densities, but at levels below the \citet{ROVILOS05} detection limits,
therefore it could be either a new source or a long lived stable
source not detected earlier because of sensitivity limitations.
Similar considerations apply to source E14. Source E10 was not
detected by \citet{ROVILOS05}, was detected in GD17A, but not detected
in GD17B.  \vskip 0.2cm

In addition to the above classification scheme, it is also interesting
to consider possible variations in 6~cm flux densities by comparing
our data with the BN022 data taken just over 2 years earlier.
Considering the thermal noise on both epochs only two sources show
$>2\sigma$ variations, namely W18 ($-$625~\uJy) and W33 ($-$505~\uJy).
Surprisingly both are class L sources.  W33 has other unusual
properties in addition to its variability and is discussed further in
Section~\ref{se:W33}.

\subsection{Source Spectra}\label{se:spectra}

To determine the spectra of our detected sources we first defined a
source position based on the highest frequency detection, then at this
pixel we determined the brightness at each of the three simultaneously
observed wavelengths (13, 6 and 3.6~cm). Tests fitting gaussians to
the four brightest sources at 3.6, 6 and 13~cm showed consistent
positions at the three wavelengths to within \about0.5 pixels in each
coordinate (one pixel being 0.25~mas). Since the source position is
defined at 3.6~cm, the maximum impact of any registration offset on
the derived spectra is set by it effect at 6~cm. At this wavelength
the map registration uncertainty amounts to less than 10\% of the beam
minor axis FWHM and will contribute only a 2\% error on the amplitude
which is negligble relative to the map noise. The resulting source
flux densities using the above procedure are given in Table~2. In some
cases the reported numbers are less than the noise or are negative,
but these measurements are still useful for constraining spectral fits
(see Section~\ref{se:fits}). An exception to the above procedure is
the resolved source W42 (see Section~\ref{se:W42}) for which the
integrated flux density at 3.6~cm was determined within a box
containing the source.

For the 18~cm flux densities a slightly different procedure was
followed because these data were not taken simultaneously with the
rest of the wavelengths and so exact position registration at the
pixel level could not be assured.  For GD17A fluxes are taken from
\citet{LONSDALE06} in the case of tabulated detections, otherwise the
map was searched at the detected high frequency position within the
18~cm beam area for anything above $3\sigma$. In no cases was any new
detection found, these sources are reported in Table~2 as $3\sigma$
upper limits. For GD17B the largest peak within an 18~cm beam area of
the high frequency position was located, in most cases large SNR
detections were found. In cases with no peak above $3\sigma$ these
upper limits are reported in Table~2.

All the measured spectra for sources in classes R, S and A are
displayed graphically in Figure~\ref{fi:spectra1}. Sources in class L
are displayed in Figure~\ref{fi:spectra2}. Taken together a wide
variety of spectral shapes are seen from inverted to flat to steep.
Overall the distribution of spectra are very similar to those of the
compact sources detected in knot A of the starburst Arp~299
\citep[see][]{NEFF04}. Further discussion and modelling of our
observed Arp~220 spectra is presented in \S~4.

%%%%%%%%%%%%%%%%%%%%%%%%%%%%%%%%%%%%%%%%%%%%%%%%%%%%%%%%%%%%%

\subsection{Source Spatial Distribution}\label{se:spatdist}

An initial impression from Figure~\ref{fi:1W} is that the 15 short
wavelength detections in the western nucleus (red circles in the top
panel) have approximately the same spatial distribution as the 18~cm
sources from \citet{LONSDALE06}. However if we only consider sources
not seen in \citet{SMITH98} (i.e. classes R, S and A) which have
rising or peaked spectra (see Figure~\ref{fi:spectra1}), these appear
to be more centrally concentrated. All seven such sources are found
within the gray rectangle in Figure~\ref{fi:1W}. In contrast amongst
the 18~cm sources of \citet{LONSDALE06} the fraction inside the
rectangle is only $17/29 = 0.58$. If the class~R, S and A sources have
the same spatial distribution there is only a 2.4\% probability of all
being within the rectangle. However this is only an {\it a posteriori}
statistic and so must be treated with caution. Future observations
plus robust statistical tests are needed to check it. Note even if
true it does not necessarily imply a difference in spectral properties
with position. These results might be explained by differences in
overall luminosity combined with the present high flux detection
threshold at short wavelength. Supporting this explanation is the fact
that the median 18~cm luminosity is a factor of two larger inside the
rectangle than outside. Furthermore a Kolmogorov-Smirnov test finds
only a 6\% probability that the 18~cm sources inside and outside are
drawn from the same flux density distribution.

\section{ANALYSIS OF SOURCE SPECTRA} \label{se:specanal}
\subsection{Cause of Low Frequency Spectral Turnovers}
\label{se:modelling}
The majority of the spectra shown in Figure~\ref{fi:spectra1} are
consistent with power law synchrotron spectra with turnovers below
some critical frequency. Such spectra are typically observed in both
SNe and SNRs \citep{{WEILER_REVIEW},ALLEN98}. In principle the low
frequency turnovers could be due to either synchrotron self absorption
(SSA) or FFA. Although the former has been successfully invoked for
weak Type~II and Type~Ib/c SNe \citep{SODERBERG05} in all successfully
modelled powerful Type~II SNe and SNRs FFA is the dominant process
\citep{WEILER_REVIEW}. This is confirmed by the fact that for those
powerful SNe and SNRs sources which are resolved the brightness
temperatures are too small for SSA to operate.

For SSA to occur at a given frequency a characteristic relationship
between source luminosity and diameter is required
\citep{CHEVALIER98}. For a given shell expansion speed this implies a
relationship between peak luminosity and rise time. Inversely for an
observed combination of these parameters there is an associated
expansion speed required for SSA to dominate \citep[see Figure~4 in
][]{CHEVALIER98}.

Sources of class~R have spectra peaking around 4~GHz and peak
luminosities of \about\E{28}~erg~\per{s}~\per{Hz}. If a rise time at
4~GHz of over 2~years is assumed (the time between GD17A detections
and BP129) then these sources lie in the region of the
luminosity--rise time diagram occupied by Type~IIn SNe. In turn, for
SSA to apply in these sources, the required expansion velocities are
$<3000$~\kms\ which is considered too slow for this kind of SN
\citep{CHEVALIER98}. The longer lived class~L sources with spectral
turnovers would require even slower expansion velocities and so are
even more unlikely to be affected by SSA. We have no idea of the age
or rise times of class~S sources.  In principle these could be rare
luminous Type~Ib/c sources with fast rise times and hence could be
affected by SSA. However it seems improbable that we detect 4 such
sources given the short time their luminosities peak.  More likely
these are class~R sources observed at an earlier evolutionary stage.
We conclude that SSA is unlikely to be a significant cause of the low
frequency cutoff in any of our sources. We will be able to test this
assumption using future observations over a wide frequency range by
examining the shape of the low frequency cutoff; unfortunately the
present sparse non-simultaneous spectral data does not allow such a
test.

\subsection{Free-free Absorption Model} \label{se:fits} In this
section we fit the observed spectra as a generic power law plus FFA.
We leave to section \ref{se:SNeSNR} the question of whether the data
are more consistent with SNe or SNRs. In SNe the FFA occurs mostly in
the wind blown bubble from the progenitor star. In contrast in SNRs
the FFA is due entirely to the foreground ionized ISM. The model
spectra we fitted are described by
\begin{equation} \label{eq:spectrum}
  S_\nu=S_{\rm sy}(\nu/1.63)^\alpha \exp\{{-\tau_{18}(\nu/1.63)^{-2.1}\}}
\end{equation}
where $\nu$ is the frequency in GHz, $S_{\rm sy}$ is the 18~cm
unobscured synchrotron flux in \uJy, $\alpha$ is the optically thin
non-thermal spectral index and $\tau_{18}$ is the optical depth at
18~cm. This model corresponds to all the FFA being foreground to the
synchrotron emission and all parts of the source being subject to the
same foreground opacity. Note the latter is not true if the FFA
absorbing medium is clumpy with clump scale sizes smaller than the
emitting source in which case more complex models will apply
\citep[][Conway et al., in preparation]{NATTA84}.

The fitting proceeded by finding for each source the global minimum of
the $\chi^2$ metric in a gridded three dimensional volume. To improve
accuracy after an initial coarse search a second search was done using
a denser grid around the coarse solution. We note that for some
sources at some wavelengths the measured flux densities were negative.
However, these values are formally our best estimates and therefore
were considered in the fitting procedure. This is a simpler and less
potentially biased procedure than fitting a mixture of measurements
and upper limits.

For the class~L sources we fitted all three free parameters of the
model to the four spectral data points provided by the three
wavelengths in BP129 plus 18~cm data from the non-simultaneous GD17B
observations (taken 0.84~year before). The use of the non simultaneous
18~cm data in the spectral fitting is justified for the class L
sources because they are known to have low variability at this
wavelength \citep{ROVILOS05}. Furthermore in all cases the GD17A and
GD17B measurements are consistent within the error bars confirming low
recent variability (compare diamonds and circles at 18~cm in
Figure~\ref{fi:spectra2}).

Solutions fitting the data within the error bars (reduced $\chi^2<
1.5$) were obtained for 6 of the 8 class~L sources, the exceptions
being W30 and W33. The optimum values for $S_{\rm sy}$, $\alpha$ and
$\tau_{18}$ are summarized in Table~\ref{ta:source_fluxes} and the
corresponding model spectra are overlayed on the data in
Figure~\ref{fi:spectra2}.  Note that in some cases the best solution
had a slightly negative opacity $\tau_{18} \approx -0.1$ which is
clearly unphysical. In those cases we set $\tau_{18}=0$ and found the
values for the other parameters which minimized $\chi^{2}$; in all
such cases the final $\chi^{2}$ was only slightly increased (by
$\lesssim 0.5$). In addition to finding the optimum values, the region
of $\tau_{18}$ and $\alpha$ around the best fit was searched finding
for each combination the optimum $S_{\rm sy}$ and recording $\chi^2$.
The area over which this $\chi^2$ increased by less than one compared
to the global minimum was used to define a 68\% confidence ellipse
around the $\tau_{18}$ and $\alpha$ solution. These confidence
ellipses are plotted in the bottom panel of Figure~\ref{fi:taualpha}
and show in most cases a significant correlation between the solutions
for the two parameters.

For the class S, R, and A sources long term monitoring data at 18~cm
is not available and we cannot assume that the GD17B data approximates
the 18~cm flux density at the time of the BP129 observations. In fact
in many cases comparison of GD17A and GD17B data suggests significant
variability. Fitting a more complex time variable model adds more
unknowns and we defer such an attempt to a later paper when we have
more data. Instead here we choose to fix $\alpha$ at a value of
$-0.72$, corresponding to the median synchrotron spectral index of the
9 well studied Type~II SNe from \citet{WEILER_REVIEW} and then fit the
remaining parameters $\tau_{18}$ and $S_{\rm sy}$ to the three
simultaneous frequency data points from BP129. We chose to do this
because fitting the full three parameter model to only three data
points resulted in large uncertainities and in some cases unphysical
optimum solutions which were clearly dominated by noise. Note that the
above fitting method can only hope to show consistency with the
assumed spectral index $\alpha$ but does not prove it.

Using the above procedure good fits passing through the three
simultaneous frequencies and consistent with a reasonable evolution of
the 18~cm flux densities are obtained for 8 of the 10 sources (see
Figure~\ref{fi:spectra1}).  The exceptions are E14 and E10. The fit
for E14 passes through the fitted data points but predicts an 18~cm
flux density very different from the consistent values seen in GD17A
and GD17B. In E10 the fit is a little too steep to fit the 13~cm data
point, on the other hand such a fit is consistent with the apparently
rapidly falling 18~cm flux density which may point to a problem with
the 13~cm data point. For the 8 well fitted sources the optimum values
of $\tau_{18}$ obtained are shown in the top panel of
Figure~\ref{fi:taualpha} including 1$\sigma$ error bars.

\subsection{Fitted Spectral Indices and Opacities}
\label{se:fitted_alphas_and_taus}

Figure~\ref{fi:taualpha} plots the fitted $\alpha$ versus $\tau_{18}$
using different symbols for the different variability classes defined
in Section~\ref{se:variability}.  This figure shows that the highest
opacities $\tau_{18}>4$ all occur in class S sources, which are those
that are newly detected.  There is only one class A source giving a
good fit and this also has quite a large $\tau_{18}=2.4$. Of the class
R or rising sources 3 out of 4 have $0.7<\tau_{18}<2.1$.  Finally
amongst the long lived class L sources 4 out of 6 have spectra
consistent with zero opacity, There appears therefore to be a rough
correlation of decreasing opacity with source type from S to R to L.
Although the exact value of $\tau_{18}$ is model dependant and
Figure~\ref{fi:taualpha} combines sources with 2 and 3 parameter fits,
the results help quantify the visual impression given from the data in
Figures~\ref{fi:spectra1} and ~\ref{fi:spectra2} of decreasing
turnover frequency along the above sequence. In
section~\ref{se:discussion} we discuss this turnover/opacity
correlation in terms of decreasing opacity with source age, consistent
with a source evolution from young SNe to more evolved SNRs.

For the class L sources there was enough data to allow us to fit for
$\alpha$ (see Section~\ref{se:fits}). This fitting gives three sources
with flat spectra ($\alpha>-0.5$), one with intermediate spectral
index ($\alpha \approx 0.6$) and finally two with steep spectra
($\alpha<-1$).  Another source falling in this last category is W7
which is not listed in Table~\ref{ta:source_fluxes} because it was not
detected at any of the BP129 wavelengths; this non detection however
implies it has a very steep spectrum ($\alpha<-1.5$).

\section{DISCUSSION} \label{se:discussion}

\subsection{Supernovae or  Supernova Remnants?} \label{se:SNeSNR}

An important question is whether the compact radio sources in Arp~220
are primarily SNe (interacting with their progenitor's CSM) or SNRs
(interacting with the denser ISM). Here we discuss the radio
properties expected from these two cases and compare with past and
present observations.

\subsubsection{Background}\label{se:background} 

Despite the extreme environment in the nuclei of Arp~220 and other
ULIRGs it is expected that wind blown bubbles with a density profile
$\rho_w(r)=\rho_o r^{-2}$ will still form around progenitor stars
\citep[][Arthur, astro-ph/0605533]{CHEVALIER01}. Hence there
should remain a distinction between the SN phase, when the blast-wave
transits this declining density wind, and the SNR phase in a constant
density ISM. Equating the ram pressure of the wind with the external
ISM pressure the expected bubble radius is
\begin{equation}
\label{eq:rb} r_{b}=0.2\dot{m}_{-4}^{0.5} v_{w1}^{0.5}p_{7}^{-0.5}\rm\quad pc
\end{equation}
where \Mdot\ is the mass loss rate in units of
\E{-4}~\Msun~\per{year}, $v_{w1}$ the wind speed in units of 10~\kms\
and $p_7$ the interstellar pressure in units of \E{7}~K~cm\tothe{-3}.
During the initial SN phase, the shock-wave moves through the stellar
wind and the optically thin radio luminosity of the synchrotron
emitting shell at a fixed radius is proportional to $\rho_{o}^k$ or
$(\Mdot/v_{w1})^{k}$ where $k$ is in the range 1.4 to 2
\citep{CHEVALIER06}. It follows that there is a very wide range in
radio luminosity for SNe depending on the wind properties (see
Figure~\ref{fi:SIGMA_D}). The most luminous RSNe with the densest
winds are optically classified as Type~IIn followed by luminous
Type~IIL/b classes \citep[for a recent review of SN types and their
radio emission see ][]{CHEVALIER06P}. Such luminous objects are
thought to be quite rare within galactic disks. For instance,
\citet{CAPPELLARO97} estimate that only 2\% of core-collapse SNe are
of the Type~IIn subclass. However there are significant uncertainties
on the fraction of SN types \citep{CHEVALIER05} and these fractions
may be strongly dependent on the number of close binaries
\citep{NOMOTO96} which could be radically different in dense ULIRG
nuclei.

During the SN phase the radio light-curves are defined by competition
between decaying synchrotron emission and decreasing FFA, giving rise
to light curve peaks which occur progressively later at longer
wavelengths \citep{WEILER_REVIEW}.

Sources are expected to enter the SNR phase, and in most cases start
to brighten, when the blast-wave reaches the edge of the bubble given
by equation~\ref{eq:rb}. The evolution just after this will be
complicated as the blast-wave transits shocked wind and possibly
(partially) collapsed HII region gas \citep{VANMARLE04,CHEVALIER04}.
Nevertheless it is expected that radio emission will peak
\citep{COWSIK84,HUANG94} at the beginning of the Sedov phase (i.e.
when the swept up mass, dominated by the dense surrounding ISM gas,
equals the ejecta mass) which occurs at radius
\begin{equation} \label{eq:sweeprad} r_s\simeq 4.1\x(M_{\rm
    ej1}/\nISM)^{1/3}\rm\quad pc
\end{equation}
where $M_{\rm ej1}$ is the ejecta mass in units of 10~\Msun\ and
\nISM\ is the ISM number density in cm\tothe{-3}. For \nISM$=$\E{4}
and $M_{\rm ej1}= 0.5$ the predicted size is $r_{s}=0.15$~pc. Once the
Sedov phase is reached the radio luminosity is expected to gradually
decrease in a way consistent with the Surface brightness--Diameter
($\Sigma-D$) relation for SNR \citep{HUANG94} \footnote{Although
  recent analysis \citep{UROSEVIC05} casts doubt on whether a
  physically significant $SIGMA-D$ relation exists for most SNR in our
  own and nearby galaxies such a correlation is confirmed for M82 and
  possibly also for SNRs in dense environments within our galaxy}.
Substituting $r_s$ into this relation we obtain a rough estimate of
the peak 3.6~cm SNR luminosity of a young SNR of
\begin{equation}
  \label{eq:L} L_{3.6}\approx 2.6\x\E{24}(M_{\rm
    ej1}/\nISM)^{-0.53}\rm\quad erg~\per{s}\per{Hz}
\end{equation}
Eventually when the internal SN gas cools to below \E{6}~K energy
losses to atomic lines become significant \citep{DRAINE91} and the
radio luminosity decays more rapidly than in the Sedov phase. If the
external density is very large (\nISM$>$10$^{5}$ cm$^{-3}$) then a
source can become radiative before it enters the Sedov phase
\citep{WHEELER80} in which case it will never reach the $\Sigma-D$
relation.

In the model of \citet{HUANG94} the SNRs all lie close to the same
$\Sigma-D$ (or Luminosity$-D$) relation because they all release
approximately the same kinetic energy (\E{51}~ergs) and a constant
fraction of this energy is injected into relativistic particles. The
more detailed modeling of \citet{BEREZHKO04} confirms that in the
early Sedov phase, for a given diameter $D$, the radio luminosity
depends only on the SN energy and is independent of the external
density. We therefore expect that all core collapse SNe would evolve
eventually to join the SNR Luminosity$-D$ relation at some diameter
defined by equation~\ref{eq:sweeprad}. Hence for an object like
SN1980K (see Figure~\ref{fi:SIGMA_D}) we can expect a dramatic
brightening when the blast-wave encounters the high density ISM and it
enters the SNR phase.

An interesting possibility is that the observed new 18~cm radio
sources in Arp~220 \citep{ROVILOS05,LONSDALE06} could be such SN/SNR
transition objects rather than SNe. Radio luminosity would start
increasing when the blast-wave reached the edge of the wind blown
bubble and would peak when the swept up ISM gas mass equalled that of
the ejecta, hence characteristic rise times would be of order
$(r_{s}-r_{b})/v_{\rm ejecta}$.  For $v_{\rm ejecta}$=\E{4}~\kms\ a
high external density of \nISM$=$\E{5}~cm\tothe{-3} and a pressure of
$p_7=1$ the rise time would take \about7~years, only somewhat longer
than SNe rise times at 18~cm. However since for an emerging SNR the
reason for the luminosity increase is quite different than for a SN (a
general increase of high energy particles and field rather than
reduced FFA) such objects are expected to show very different
multi-wavelength light curves.  According to the model of
\citet{BEREZHKO04} the radio spectrum is expected to keep
approximately the same shape during the rise phase and hence we would
expect the multi-wavelength light-curves to show similar behavior, in
contrast to the characteristic delay between wavelengths observed in
SNe.

At the other end of the SN luminosity range (see
Figure~\ref{fi:SIGMA_D}) for objects like SN1986J and SN1979C the
observational data are consistent with a fairly smooth luminosity
transition between the SN and SNR phases. This might be expected
because the progenitors of such powerful RSNe have the densest winds
and therefore the smallest density contrasts between their terminating
wind and the external medium.

\subsubsection{Previous  Observations}
\label{se:prevobs}

In their discovery paper \citet{SMITH98} proposed a SN based model for
the compact sources in Arp~220 in which all were due to extremely
luminous Type~IIn SNe. The number of observed sources and the total
FIR luminosity were found to be consistent if all these SNe were as
radio luminous as the Type~IIn supernova SN1986J. \citet{CHEVALIER01}
argued that this model was unlikely because such luminous objects
comprise only a small fraction of core-collapse SNe. Observational
doubts on the \citet{SMITH98} model were cast by the 18~cm light curve
monitoring of \citet{ROVILOS05} which did not show the decay in flux
expected from SN1986J like sources. Consequently both
\citet{ROVILOS05} and \citet{LONSDALE06} have suggested that most
sources in Arp~220 are instead higher luminosity versions of the SNRs
in M82.

The above SNR hypothesis seems plausible if we consider the expected
and observed flux densities. In equation~4 we estimated the maximum
luminosity of a SNR at wavelength 3.6~cm as a function of ISM density.
Assuming a spectral index of $-0.7$ and scaling to the distance to
Arp~220 this predicts 18~cm flux densities of $1.2\x(M_{\rm
  ej1}/\nISM)^{-0.53}$~\uJy. For the median 18~cm source flux density
in the western nucleus of 207~\uJy\ \citep{LONSDALE06} and assuming
$M_{\rm ej1}=0.5$ we obtain an estimated ISM density of
\nISM=8300~cm\tothe{-3}\ which is consistent with the mean molecular
density of 1.5\x\E{4}~cm\tothe{-3}\ estimated by \citet{SCOVILLE97}.

In addition to the slow evolution of the brightest sources
\citet{ROVILOS05} and \citet{LONSDALE06} also detected some \emph{new}
weak 18~cm sources. These papers propose that these new sources are
SNe, implying a mixed SNe and SNRs population. Supporting the SN
interpretation for the new sources it was found that if their rate of
appearance was taken as the SN rate $\nu_{\rm SN}$ and a standard IMF
assumed then the predicted SFR was consistent with that estimated from
the FIR luminosity \citep{LONSDALE06}. A possible difficulty however
is the assumption that all core-collapse SNe give rise to RSNe
luminous enough to be detected. As described in
Section~\ref{se:background} such luminous radio sources are thought to
be rare. One possible explanation is that the new radio sources are
instead SN/SNR transition objects (see Section \ref{se:background}).
This and other possible explanations for reconciling the rate of new
radio sources and the SFR are discussed further in
Section~\ref{se:SFR}.

\subsubsection{Comparison of New Data and SNe Models}
\label{se:SNe}

The standard model for the radio emission from powerful RSNe comprises
power law emission from a synchrotron emitting shell observed through
FFA gas from the ionized CSM \citep{WEILER_REVIEW}. This absorption
decreases rapidly with time as the blast-wave moves outward. In
Section~\ref{se:fits} we fitted our observed source spectra around the
epoch BP129 with such a power law plus FFA model and the estimated
spectral indices and opacities were plotted in
Figure~\ref{fi:taualpha}. This plot shows the expected correlation
between source age and opacity predicted by the RSN standard model.
All the recently detected (class S and R) sources have high opacity,
in contrast to those known to be old (class L) which have lower
opacity. This strongly suggests a SN origin for the class S and R
sources.

To further check whether the standard SN model applies to our sources
Figure~\ref{fi:TAU_S} plots the fitted $\tau_{18}$ versus flux density
at 3.6~cm wavelength ($S_{3.6}$). Superimposed are loci traced by
fitted models to four powerful Type~II SNe with well sampled
multi-frequency light curves \citep{WEILER_REVIEW} scaled to the
distance of Arp~220.  $S_{3.6}$ and $\tau_{18}$ are natural
coordinates to plot our data, since for sources beyond the 3.6~cm peak
luminosity they decouple the intrinsic strength of the synchrotron
emitting shell from the foreground absorption. In this diagram the
effect of small foreground ISM opacities, which are insufficient to
affect the 3.6~cm flux densities, is to move the SN tracks to the
right by adding a constant $\tau_{\rm ISM}$.

Figure~\ref{fi:TAU_S} shows that for all the well fitted sources in
classes S, R and A both the opacities and luminosities are in the
range defined between the most luminous detected Type~IIL source
SN1979C and the Type~IIn source SN1986J. The best fit is in fact to
the intermediate luminosity Type~IIn SN1978K. However, if there is a
foreground ISM opacity at 18~cm of $\tau_{18}\approx 1$ then SN1986J
instead provides the best fit. In both cases the required ages as of
the BP129 observational epoch (January 2006) would be in the range 1
to 7 years. This result suggests that sources in these three
variability classes are powerful SNe.

In contrast it seems very difficult to explain class L sources using
the standard RSN model. Three of them (W10, W17 and W42; the three
relatively flat spectrum sources shown on the top row of
Figure~\ref{fi:spectra2}) have $\tau_{18}<0.3$ and 3.6~cm fluxes over
250~\uJy. These could be consistent with objects like SN1986J but only
if they had ages $<8$ years, but this is less than their minimum ages
($>11$~years) ruling out a standard SN origin. Sources W18 and W39
could be SNe, however both sources have steep spectra (see middle row
of Figure~\ref{fi:spectra2}) and the weakness of their 3.6~cm flux
could be ascribed to this.  Finally source W8 is in a part of the
diagram that could be reached by SN models but again the required age
is much smaller than is observed.

\subsubsection{Comparison of New Data With SNR Models}
\label{se:SNR}

In the previous section it was shown that class S, R an A sources are
consistent with being RSNe. In contrast for the older class L sources
known since \citet{SMITH98} the situation is less clear. These sources
have been monitored at 18~cm by \citet{ROVILOS05} and Thrall et al.
(in preparation) and do not show the expected flux density decay for
SNe. Therefore we should consider the possibility that these are SNRs
interacting directly with the dense ISM.

To study this possibility further it is interesting to compare our
class-L sources with the radio SNR in M82 which have been well studied
in a series of papers beginning with the work of \citet{KRON85},
\citet{BARTEL87}, \citet{UA93} and \citet{MUX94}. In
Figure~\ref{fi:SIGMA_D}\ we plot the 3.6~cm flux density versus
diameter for the six well fitted class L sources. This plot also shows
the integrated fluxes of the SNRs in M82 scaled down to the distance
of Arp~220. The diagonal line represents the empirical relation
% $\Sigma=6.6\times D^{-3.6\pm0.1}$ 
between surface brightness $\Sigma$ and diameter $D$ for Galactic, LMC
and M82 SNRs compiled by \citet{HUANG94} converted to luminosity and
then to flux density at the distance of Arp~220. As described in
Section~\ref{se:background} SNRs in the Sedov phase of their evolution
are expected to lie close to this line.

The location in the diagram of the only resolved source W42 is well
above the empirical relation (see Section~\ref{se:W42}). For the
remaining sources we only have upper limits on size as indicated by
the horizontal arrows. However if we assume these follow the SNR
luminosity--size relation we can estimate their sizes. It is clear
from the M82 data that there is at least a factor of 3 dispersion
about the relation. Because we expect that our detections are biased
toward the most luminous sources at a given diameter, we estimate
diameters in the range 0.17 to 0.4~pc with median 0.2~pc. These
diameters have to be greater than $2r_{b}$ and $2r_{s}$ given in
Equations~\ref{eq:rb} and \ref{eq:sweeprad}, setting lower limits to
respectively the ISM pressure and density.

Taking a typical radius limit of 0.1~pc then for a Type~IIn progenitor
with $\Mdot=1$ and $v_{\rm w1}=1$ the required ISM pressure is $P_{\rm
  ISM}>$4\x\E{7}~K~cm\tothe{-3}. For more typical Type~II progenitors
with lower mass loss rates the pressure limit will be reduced
proportionally to $\Mdot$ (but this may be partially offset by higher
wind velocity). This pressure is similar to that found by modelling
the spectral energy distribution of Arp~220 of \E{7}~K~cm\tothe{-3}
\citep{DOPITA05}. The somewhat higher pressure we obtain can be
explained as a selection effect in which we preferentially detect SNR
in the highest density and pressure regions (see equation~\ref{eq:L}).

Turning to estimates of ISM density, if in equation~\ref{eq:rb} the
ejecta mass is taken as $M_{\rm ej1}=0.5$ then
\nISM$>$3\x\E{4}~cm\tothe{-3}. Although this is only a lower limit on
density we expect our detected sources to be embedded in environments
with number densities close to this value. This is because given our
limited sensitivity we are probably only detecting the most luminous
SNRs that exist within Arp~220. Given the expected high $\nu_{\rm SN}$
of $4\pm2$~\per{year} \citep{LONSDALE06} these brightest and youngest
SNRs in high density environments are probably being seen only years
or decades after reaching their radio maximum and hence have radii
close to $r_{s}$. The estimated ISM densities for the short wavelength
detected sources are larger by a factor of a few compared to those
estimated for the weaker 18~cm sources (see \ref{se:prevobs}).  The
density estimates are consistent with the mean density estimate by
\citet{SCOVILLE97} of 1.4\x\E{4}~cm\tothe{-3} based on CO(1-0)
observations. Recent CO(3-2) observations \citep{NARAYANAN05} in
ULIRGs including Arp~220 confirm, via the tight observed correlation
with FIR luminosity, that most massive star formation occurs in
environments with densities $>$1.5\x\E{4}~cm\tothe{-3}.  Note that the
predicted ISM densities for our bright SNR sources are just below the
boundary (\nISM$>$\E{5}~cm\tothe{-3}) at which they would become
radiative before reaching their Sedov phase \citep{WHEELER80}.

\subsubsection{Probable Nature of the Arp~220 Compact Sources}
\label{se:probnat}

Based on comparison with data from well monitored radio SN there
appears to be good evidence that the four class R sources, with rising
flux density at 18~cm wavelength are from Type~IIn RSN. Likewise the
three class S sources which are so far detected only at high frequency
have spectra consistent with being even younger SNe where the FFA is
still optically thick at 18~cm. Since however we have detected these
sources at only one epoch we cannot yet rule out the possibility that
they are steep spectrum SNe or SNRs with extreme foreground ISM
absorption. Only further monitoring observations will be able to
definitely decide the issue. Similar considerations apply to the three
class A sources (W15, E10, E14). The source W15 is well fitted by a
standard spectral index $\alpha = -0.72$ and moderate FFA.  This
source shows no variability at 18~cm between GD17A and GD17B and is
most likely a stable class L source which is just too weak to have
been detected by \citet{SMITH98} and \citet{ROVILOS05}. Source E14
shows inconsistency between the measured 13~cm flux and the two
earlier 18~cm measurements and is hard to classify. Source E10 which
shows 18~cm variability is possibly a SN but as yet we do not have a
successful spectral fit.
  
It should be noted that so far there is no need to resort to models of
SN/SNR transition objects (see Section \ref{se:prevobs}). to explain
the properties of class R or S sources. These transition objects if
they exist are expected to show different multi-frequency lighcurves
which change in unison (see Section~\ref{se:background}) rather than
having the characteristic delay towards longer wavelengths of SN
models. Future multi-frequency monitoring should be be able to give a
conclusive result on whether any such objects exist.

For the eight detected long lived class L sources which were in the
sample of \citet{SMITH98} the analysis in Sections~\ref{se:SNe} and
\ref{se:SNR} shows that although a couple could be decades old SNe
most have difficulties with this interpretation. The main problem, as
already noted by \citet{ROVILOS05}, is the high luminosity and
stability of the observed light curves given their age. A striking
thing about these sources is the diversity of their spectra (see
Figure~\ref{fi:spectra2}).  One source (W18) has a normal synchrotron
spectral index $\alpha=-0.62$ and a peaked spectrum caused by moderate
absorption and is similar in form to the several of the none class L
sources shown in Figure~\ref{fi:spectra1}. Three sources (W10, W17 and
W42) have flat ($\alpha >-0.5$) spectra, while another two sources (W8
and W39) have steep ($\alpha <-1.9$) synchrotron spectra. Finally W33
has a complex spectrum that cannot be fitted by standard models. For
all the above class L sources (except perhaps W33) it was argued (see
Section~\ref{se:SNR}) that a plausible origin was from SNRs in dense
ISM environments. Still the diversity of their spectra is a puzzle. We
discuss these sources in more detail in \ref{se:particular}.

\subsection{Discussion on Particular Source Classes}
\label{se:particular}

In this section we discuss individual sources classes with unusual
properties. All of these sources are found amongst our class L
sources.

\subsubsection{Steep Spectrum Sources}

Two of the class L sources (W8 and W39) have fitted synchrotron power
law spectra with $\alpha<-1.2$.  In addition W7 which is not detected
at any wavelength shortward of 18~cm must have $\alpha<-1.5$.  These
sources (see the middle row of Figure~\ref{fi:spectra2}) are amongst
the brightest at 18~cm in \citet{LONSDALE06} and were already detected
at 18~cm by \citet{SMITH98} in observations from 1994, implying ages
of at least 11 years. We argued in Section ~\ref{se:probnat} that
these and other class L sources were SNRs.  In our galaxy SNRs with
such extreme spectral indices appear to be highly unusual. In the
catalog of galactic SNR compiled by Green\footnote{Green D. A., 2006,
  `A Catalogue of Galactic Supernova Remnants (2006 April version)',
  Astrophysics Group, Cavendish Laboratory, Cambridge, United Kingdom
  (available at {http://www.mrao.cam.ac.uk/surveys/snrs/}).}
comprising 265 objects there are none with $\alpha<-1$.  However in
M82 there is one source (42.53+61.9) catalogued with $\alpha=-1.84$
\citep{McDONALD02}.

Possible physical origins for the steep spectrum sources in Arp~220
are unclear. The SNR models of \citet{BEREZHKO04} predict in the early
Sedov phase spectral breaks which lie well above the radio regime
Adiabatic expansion caused by a SNR evolving into a low density region
(perhaps escaping its parent molecular cloud) could shift the break to
lower frequencies but only at the cost of dramatically dimming the
source which is clearly inconsistent with the fact that these sources
are amongst the brightest observed at 18~cm. \citet{ASVAROV06} note
that it is difficult under the standard theory of diffusive shock
acceleration in SNR to get an injected electron energy distribution
giving spectral indices $\alpha<-0.6$, though a suggested possibility
is that such spectra could arise if a SN exploded into a pre-existing
cavity with increasing density versus radius.

\subsubsection{ Flat Spectrum Low Opacity Sources}
\label{se:flat}

Three of the class L sources (W10, W17 and W42) show relatively flat
($\alpha>-0.3$) spectra with no sign of low frequency turnovers (see
top row of Figure~\ref{fi:spectra2}). The fitted spectral index and
the straightness of the spectra will have to be confirmed by future
observations. A cautionary tale is the case of 44.01+596 in M82 which
was initially considered an AGN candidate by \citet{WILLS97} on the
basis of its apparently slightly inverted spectrum between 18 and
3.6~cm \citep[see Figure 11 in][]{WILLS97}.  Further observations by
\citet{ALLEN98} extending to shorter wavelengths showed an overall
spectrum that could be fitted by a conventional $\alpha \approx -0.5$
power law plus FFA.

Despite the above caveat the spectra for these three sources are
sufficiently striking as to suggest a possible distinct class of
source. In M82 a few such similar sources with $\alpha>-0.3$ exist
\citep{McDONALD02}. Interestingly these sources appear from MERLIN and
VLBI imaging to show more complex spatial structures than the shell
like structures seen amongst the steeper spectrum sources
\citep{McDONALD02}. In the starburst Arp~299 the spectra of compact
sources A2, A3 and A4 \citep{NEFF04} are also consistent with being
fairly flat spectrum. Amongst galactic SNRs in the catalog maintained
by Green approximately 10\% of the shell type remnants have
$\alpha<-0.4$.  There is some evidence that these flat spectrum
sources are preferentially found in dense environments; for instance
the two prototypical galactic SNRs within molecular clouds, W44 (not
to be confused with the source of the same name in Arp~220) and IC443,
have $\alpha=-0.39$ and $\alpha=-0.36$ respectively
\citep{OSTROWSKI99}.  Theoretically spectral indices $>-0.5$ can be
explained by second order Fermi acceleration \citep{OSTROWSKI99}
amongst other mechanisms \citep{ASVAROV06}.
 
A more exotic possibility is that these flat spectrum objects are
plerions powered by extraction of spin energy from a central neutron
star. Recent VLBI observations of the archetypical radio luminous
Type~IIn SN1986J \citep{BIET04} show a compact high frequency source
which has appeared 16 years after the initial explosion. After the
emergence of this component the overall spectrum is observed to be
flat shortward of 13~cm. If placed at the distance of Arp~220 the
observed flat spectrum source in SN1986J would be only 74~\uJy\ in
flux density, however the central source may still be rising in
luminosity. Theoretically \citep{BIET04} the extraction of energy from
a central pulsar could produce a source up to 5 times as luminous as
presently observed in SN1986J and hence be comparable in luminosity to
the flat spectrum sources we see in Arp~220. Finally there is the
possibility, which we mention for completeness, that one or more of
the flat spectrum sources might be due to an AGN or an accreting
intermediate mass black hole \citep{WROBEL06}.

\subsubsection{W42: Resolved Flat Spectrum Source} \label{se:W42}

W42 is the only source showing evidence for resolution at 3.6~cm. A
deconvolved FWHM of 2.3~mas is obtained using a Gaussian fit which
corresponds to a linear size of 0.86~pc. Although we cannot rule out
the possibility that it could be two or more sources close together we
here assume a single source. W42 is one of the three sources showing
relatively flat spectra (see previous section). The source was
detected in the original discovery observations of \citet{SMITH98} and
hence its minimum age is 11 years implying an upper limit of
40000~\kms\ for its mean expansion velocity. However monitoring at
18~cm \citep[see Figure~3 in][]{ROVILOS05} shows a stable light curve.
This stability persists to the latest 18~cm measurements (see
Table~\ref{ta:source_fluxes}, Thrall et al., in preparation) and
suggests that the source is significantly older.

Estimates of minimum energy in relativistic particles and fields with
W42 can be made using the standard equipartition argument
\citep[see][]{LONGAIRV2}. Recently problems with this standard
derivation have been pointed out by \citet{BECK05} who provide revised
formulae to calculate minimum fields and energies for cases with
$\alpha<-0.5$.  Since for spectral indices close to $-0.5$ differences
to the classical result are relatively small we instead use the
version of the classical formula from \citet{BECK05}, which is
applicable for any $\alpha$, to calculate the minimum energy
requirement for W42.

Along the position angle corresponding to the beam minor axis the FWHM
of W42 is about twice as big as the beam FWHM (see
Figure~\ref{fi:contours}-Left) being consistent with a thin shell of
diameter 2.3~mas or 0.86~pc. If we assume that this radio emitting
shell has a ratio of inner to outer radius of $0.8$, similar to the
SNR shells imaged in M82 \citep[see for
instance][]{KRON85,UA93,MUX94}, then the volume filling factor
$\phi=0.51$.  We take the upper limit of integration of the spectrum
to be 8.6~GHz which is the highest frequency for which we so far
detect emission. A critical parameter in the classical formula is the
ratio of total energies in protons to electrons $K$. For a source with
a flat spectral index this ratio will be similar to the relative
number densities of protons and electrons at fixed energy which in
SNe/SNRs is probably in the range 40 to 100 \citep[see][]{BECK05}; we
adopt $K=100$.  Using the above parameters the resulting minimum total
energy in W42 is $\sim$ 2\x\E{50}~ergs.  This value is a weak function
of the assumed $\alpha$ and upper frequency cutoff. The energy scales
as $(1+K)^{4/7} \phi^{3/7}$ so the energy requirement could be
decreased if we are seeing radio emission from low volume filling
factor filaments or if the overall structure is not spherical but
bipolar or elongated.

\citet{HUANG94} estimates for young SNRs entering the Sedov phase, in
which a large fraction of the initial kinetic energy has been
converted to thermal energy, the sum of energy in relativistic
particles and fields is 2\% of this thermal energy. If W42 has the
same energy conversion efficiency then the total kinetic energy
released by its progenitor SN is \E{52}~ergs, which is ten times
larger than the canonical value of \E{51}~ergs expected for
core-collapse SNe. There is however increasing evidence that
\emph{hypernova} remnants created from such energetic explosions do
occur \citep[see for instance][]{UROSEVIC05}. Long period Gamma Ray
Bursts are now thought to come from a subclass of Type~Ibc SN
\citep{WOOSLEY06} with total kinetic energies \about 5\x\E{51}~ergs.
Such an energetic hypernova origin for W42 would be consistent with
its high luminosity relative to the SNR Luminosity$-D$ relation (see
Figure~\ref{fi:SIGMA_D}) which according to \citet{BEREZHKO04} depends
only on its its total kinetic energy.

It is interesting to compare W42 with the two brightest 3.6~cm sources
in M82 (see Figure~\ref{fi:SIGMA_D}) both of which have similar
diameters but luminosities 10 times smaller. The source 41.95+575 is
the brightest, most compact source in M82 and appears as a bipolar
structure of \about0.5~pc along its major axis \citep{BESWICK06}. Like
W42 it also lies well above the Luminosity$-D$ relation but is less
extreme in this regard. It also has some other differences including a
significant decrease in flux density of 7.1\%~\per{year}
\citep{BESWICK06}, an estimated expansion velocity of only
\about1500--2000~\kms\ and a conventional peaked radio spectrum. The
next most luminous source in M82 is 44.01+596, which shows a shell
structure of diameter 0.79~pc \citep{HUANG94,McDONALD01}. This is
probably the closest analog to W42; as noted in the first paragraph of
this section this object was initially thought to have an unusual flat
spectrum but subsequent observations over a wider frequency range
showed a conventional peaked spectrum source.

\subsubsection{Complex Spectrum Source W33}\label{se:W33}

\begin{figure*}[t]
  \centering
\includegraphics[width=0.5\hsize]{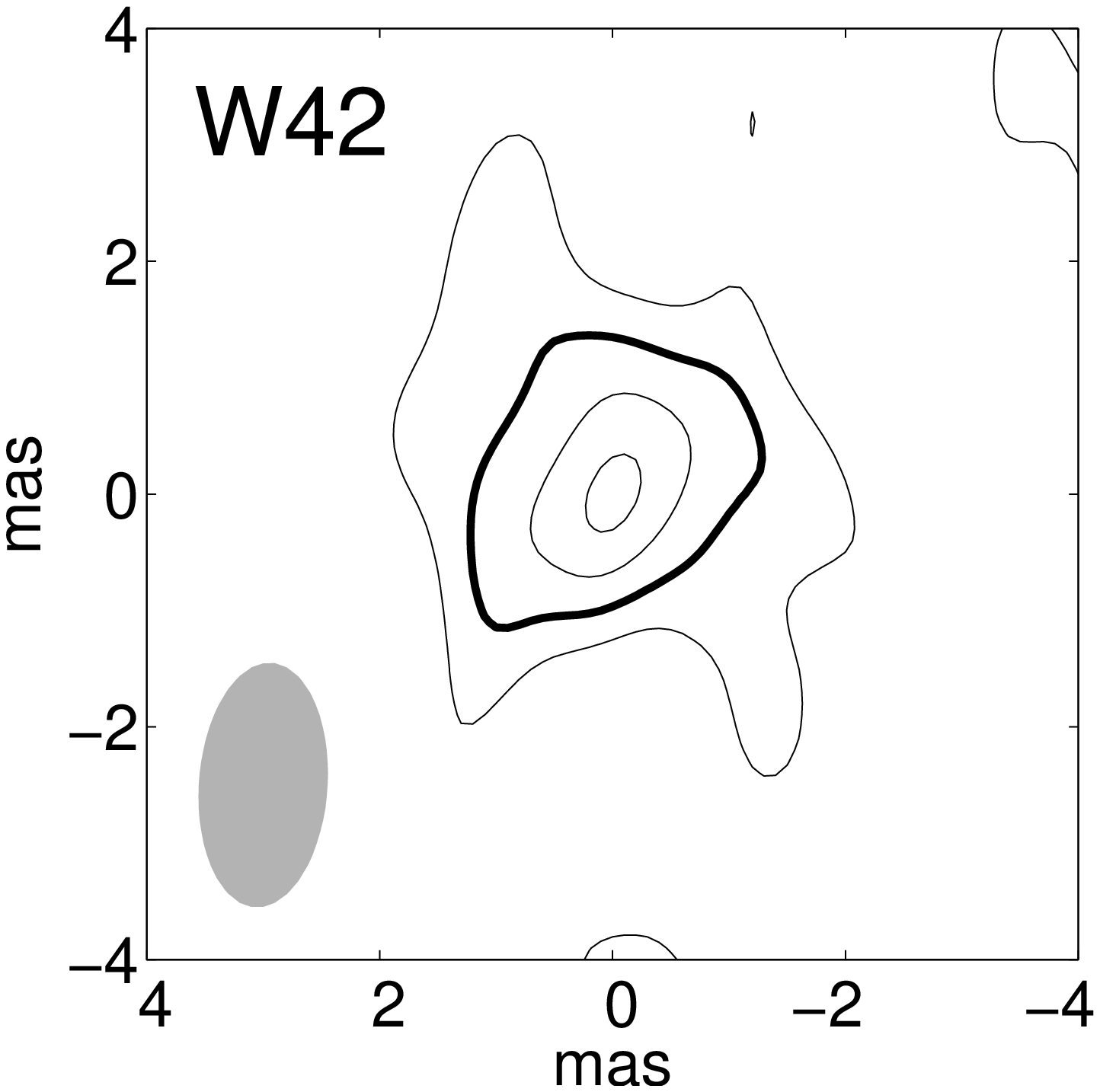}\hfil
\includegraphics[width=0.37\hsize]{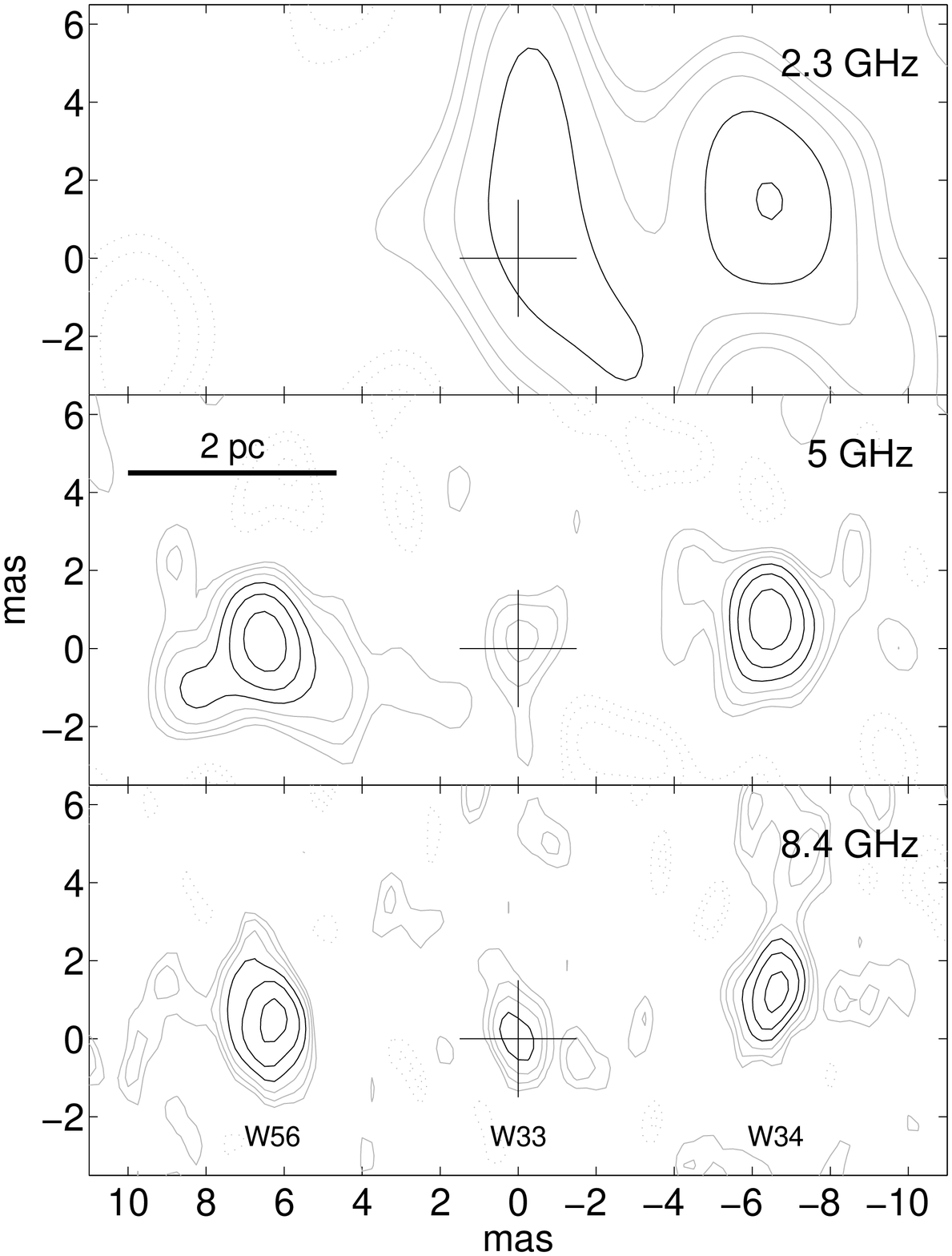}\hfil
\caption{Detailed structure of source W42 and of the W33 region.  {\bf
    Left: } Contour map of W42 produced from the 3.6~cm wavelength
  uniformly weighted image. Contour levels are $-$25, 25, 50, 75 and
  95\% of the peak (see table \ref{ta:source_fluxes}). The 50\%
  contour is drawn with a thick line and the $-$25\% ones with dashed
  (none visible).  The 50\% contour appears elongated in the NE--SW
  direction with respect to the beam (shown in the bottom left corner)
  indicating source resolution. {\bf Right: } Contour maps of the W33
  region at 13, 6 and 3.6~cm. Contour levels are
  $\ldots,-2,-\sqrt{2},\sqrt{2},2\ldots$ multiplied by the
  corresponding map noise (see Table~\ref{ta:epochs}). Contour levels
  below the detection threshold at every frequency are drawn in gray.
  Negative contours are drawn with dotted lines. The axes are in mas
  relative to the peak in the 3.6~cm map whose position is indicated
  by a cross in all three panels (see
  Table~\ref{ta:source_fluxes})}\label{fi:contours}
\end{figure*}

\begin{figure*}
\centering
\includegraphics[width=1\hsize]{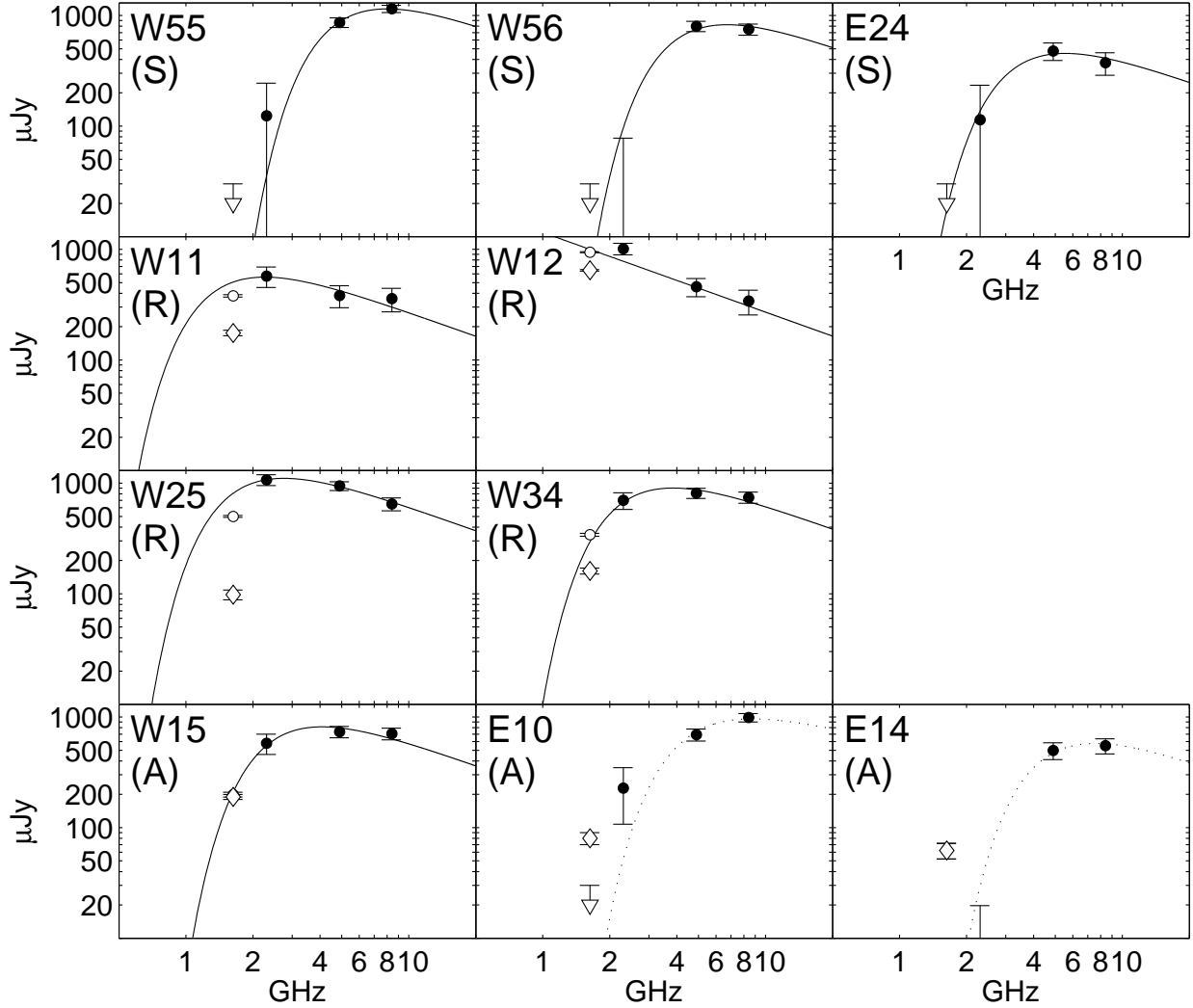}
\caption{Spectra and models of short wavelength detected sources not
  in the original 18~cm sample of \citet{SMITH98}; these sources are
  either new since 1995, weak at long wavelengths or both. The
  variability subclass to which each source belongs (R, S and A, see
  Section~\ref{se:variability} for definition) is indicated in the top
  left corner of each spectrum. The data at 13~cm (2.3~GHz), 6~cm
  (5~GHz), 3.6~cm (8.4~GHz) were taken from simultaneous observations
  in BP129 and are shown as filled circles.  For those points error
  bars are plotted at $\pm1\sigma$ about the measured value given in
  Table~\ref{ta:source_fluxes}. In cases where the measurement gives a
  negative flux density only the upper error bar is plotted. The 18~cm
  data comes from the earlier epochs of GD17A and GD17B, shown as
  diamonds and open circles respectively. In the case of detections at
  this wavelength error bars are plotted at $\pm\sigma$. If a source
  is undetected at one or more of the 18~cm epochs an upper limit is
  plotted.  The curves shown are the best fits of the model spectrum
  described in section~\ref{se:modelling} obtained by fitting only to
  the simultaneous 13, 6 and 3.6~cm data from the BP129 epoch. For
  sources E10 and E14 the best fits are shown with dotted lines
  indicating that these models were unable to fit all the data within
  the error bars.}
\label{fi:spectra1}
\end{figure*}

\begin{figure*}
\centering
\includegraphics[width=1.00\hsize]{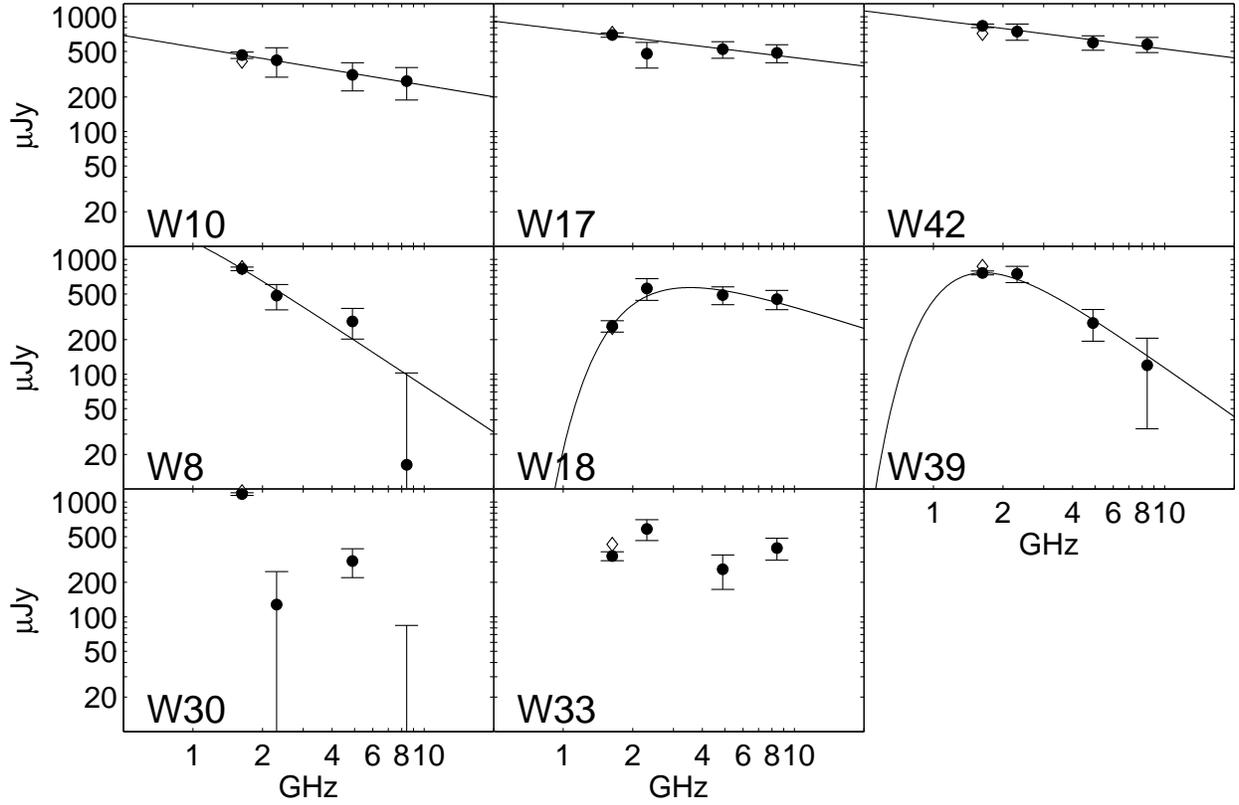}
\caption{Spectra and models for the short wavelength sources detected
  in the original 18~cm sample of \citet{SMITH98} (class L sources,
  see Section~\ref{se:variability}). In all cases error bars are
  plotted at $\pm1\sigma$ about the measured value given in
  Table~\ref{ta:source_fluxes}. When the measured flux density is
  negative only the upper error bar is plotted. The first row shows
  three sources with relatively flat spectra. The middle row shows the
  remaining sources that are well fitted by a power law synchrotron
  plus FFA model (see Section \ref{se:modelling}).  Finally the last
  row shows two sources for which good fits could not be obtained. The
  data at 13~cm (2.3~GHz), 6~cm (5~GHz), 3.6~cm (8.4~GHz) were taken
  from simultaneous observations in BP129. The 18~cm data from earlier
  epochs GD17A and GD17B are shown as filled circles and open diamonds
  respectively (in some cases these data points overlie each other).
  Error bars are shown at $\pm\sigma$ (see Table~\ref{ta:epochs}) The
  curves shown are the best fits of the synchrotron plus FFA model
  spectrum described in section~\ref{se:modelling} as fitted to the
  data at all four wavelengths. For sources W30 and W33 no reasonable
  fits could be obtained and therefore no model is plotted.}
\label{fi:spectra2}
\end{figure*}

\begin{figure}
  \centering \includegraphics[width=0.7\hsize]{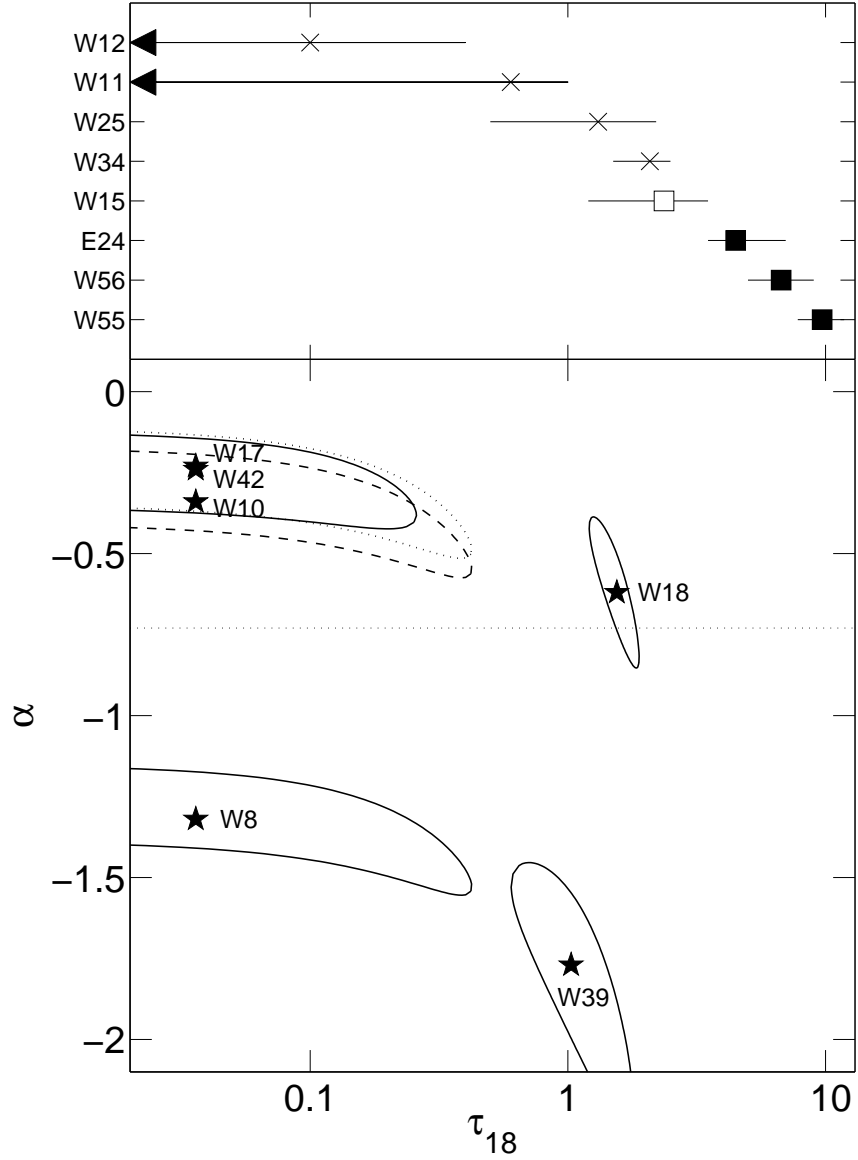}
  \caption{Summary of spectral model fitting results for the power law
    plus free-free absorption model presented in
    Section~\ref{se:fits}. {\bf Bottom: } The stars indicate the best
    fit values of $\tau_{18}$ and $\alpha$ for the six class L sources
    with three parameters fits to four wavelength spectral data. Note
    that for sources W8, W10, W17 and W42 the optimum value of
    $\tau_{18}$ is estimated as zero which cannot be plotted on a on a
    logarithmic scale and therefore these points are plotted at an
    arbitrary small value of $\tau_{18}$. The surrounding contour
    around each plotted point delineates the region in which the
    reduced $\chi^{2}$ increases by less than one compared to the best
    fit, indicating the 68\% confidence region. (The contours for W17
    and W10 are drawn in dotted and dashed line respectively). The
    dotted horizontal line is plotted at the value of spectral index
    assumed for the two parameter fits shown in the top panel. {\bf
      Top: } Shows fitted $\tau_{18}$ for those sources fitted using
    only the three wavelengths from BP129 and fixing synchrotron
    spectral index $\alpha=-0.72$ (see section~\ref{se:fits} for
    details). Lines indicate $1\sigma$ error bars.  Sources from
    variability classes S, R, and A are indicated as filled squares,
    crosses, and empty squares respectively.}
  \label{fi:taualpha}
\end{figure}

\begin{figure}[t!]  \centering
  \includegraphics[width=1\hsize]{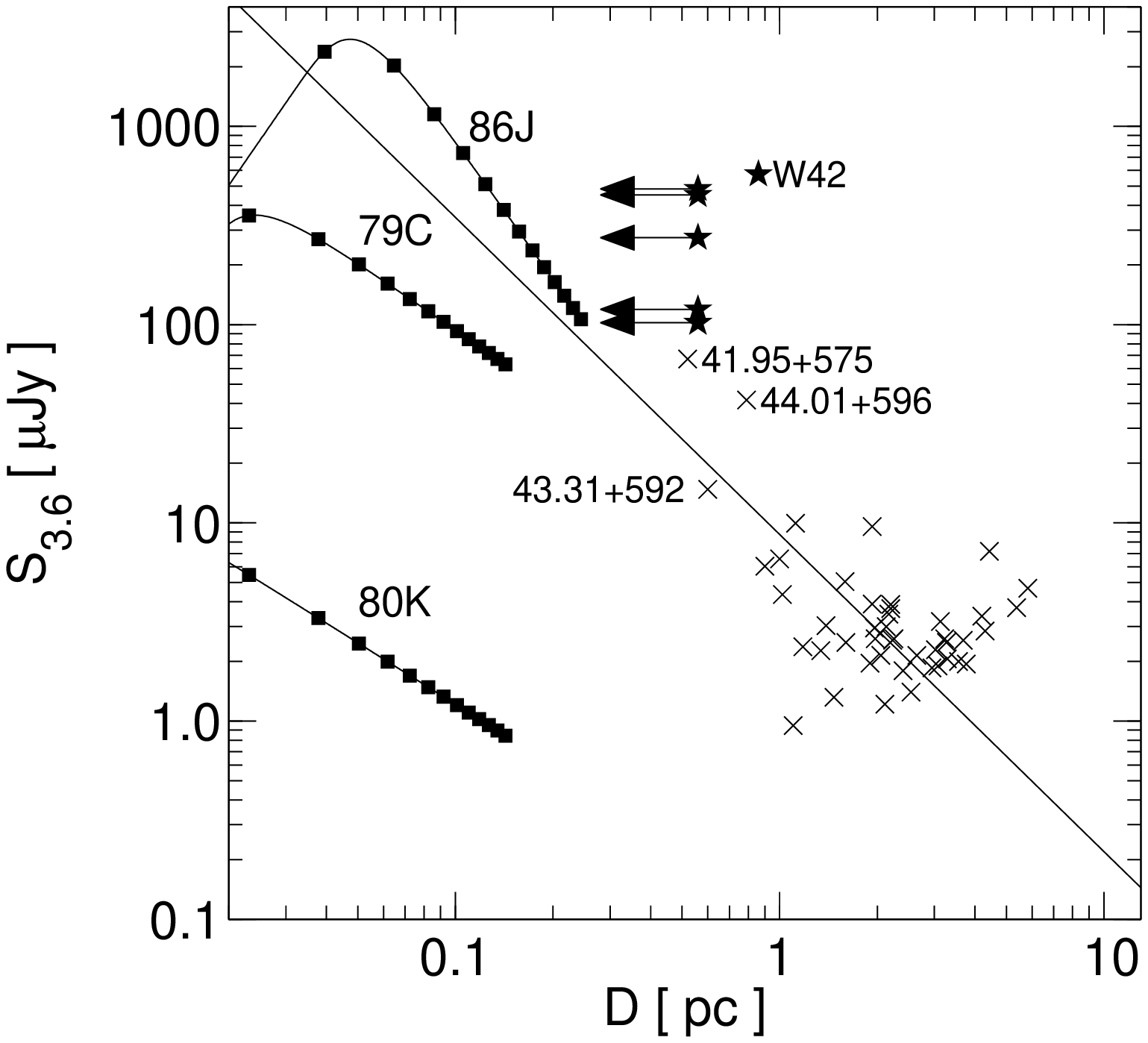} \caption{Illustration of the
    expected 3.6~cm flux density and sizes of observed SNe and SNRs if
    placed at the distance of Arp~220. The SN tracks for SN1986J
    (Type~IIn), SN1979C (Type~IIL) and SN1980K (Type~IIL) were
    produced using the light curve fits given in \citet{WEILER_REVIEW}
    combined with the deceleration parameter from \citet{BIET02} for
    SN1986J and assuming free expansion at \E{4}~\kms\ for both
    SN1980K and 1979C. The square markers along each track indicate
    time evolution and are 1~year apart. Crosses indicate the location
    of 45 SNRs in M82 from \citet{HUANG94} with their fluxes scaled
    down to the distance of Arp~220. The diagonal line is derived from
    the empirical relation between SNR surface brightness and diameter
    ($\Sigma-D$) fitted by \citet{HUANG94} converted to flux density
    at the distance of Arp~220. Stars show data for seven class L
    sources (W33 is not shown). Horizontal arrows indicate upper
    limits in size for the unresolved sources.}  \label{fi:SIGMA_D}
\end{figure}

\begin{figure}[t!]  \centering
  \includegraphics[width=1\hsize]{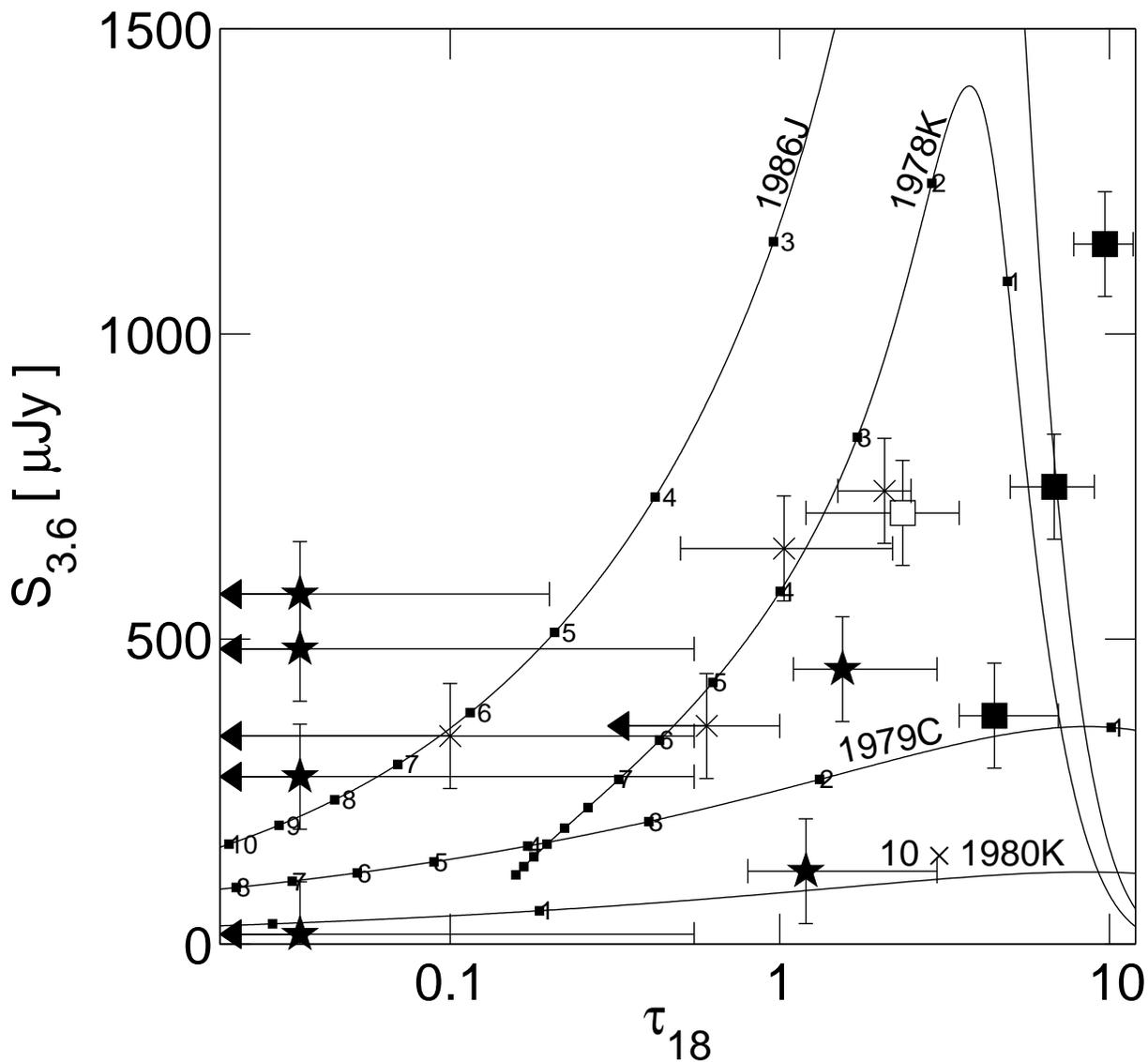}\hfil \caption{Comparison 
    of observed flux density at 3.6~cm ($S_{3.6}$) and fitted opacity
    properties ($\tau_{18}$) of the Arp~220 sources to those of well
    studied RSNe. Arp~220 sources are shown using the same markers as
    in Figure~\ref{fi:taualpha}. Lines are the loci traced by the best
    light curve fit \citep{WEILER_REVIEW} to the observations of
    SN1986J, SN1978K (both Type~IIn), SN1979C and SN1980K (both
    Type~IIL). Note that the curve for SN1980K has been amplified by a
    factor of 10 for better visualization. Markers on the tracks are
    labeled with the time after shock break-out in years.}
  \label{fi:TAU_S} \end{figure}

The region around source W33 has a complex structure as shown by
Figure~\ref{fi:contours}-Right. Figure~\ref{fi:spectra2} shows the
spectrum evaluated at the position of W33 itself, as marked by a cross
in Figure~\ref{fi:contours}-Right.  Note in this particular spectrum
the vertical axis is the brightness in \uJy~\per{beam}. Because of its
complex nature a power law plus FFA model did not give a good fit to
this spectrum (see Section~\ref{se:fitted_alphas_and_taus}). Another
unusual property of W33 is that at 6~cm the source has shown a rapid
decrease in flux density of 40\% in just over 2 years (the time
difference between the BN022 and BP129 epochs). Being a class L
source, it is known to have a long lifetime. The light-curve at 18~cm
shows no overall trend but does have significant variability between
epochs \citep{ROVILOS05}.

The 13~cm image in Figure~\ref{fi:contours} shows evidence for a
possible extension to the north of W33 which appears to be significant
with respect to the noise. However no such extension is seen in the
sensitive 18~cm image made from data taken less than a year before
(GD17B). This fact plus similar weaker extensions to the source
immediately to the west (W34) suggests that the W33 extension may
simply be an imaging artifact. We cannot however exclude the
possibility that it could be due to an adjacent Type~Ib/c SN which is
expected to evolve on timescales of $<100$ days but can in some cases
reach or exceed the luminosity of Type IIn's
\citep[see][astro-ph/0607422]{CHEVALIER06}. It is interesting that the
region around W33 lies close to one end of a patch of high brightness
18~cm emission (Thrall et al., in preparation) which may be either
high brightness diffuse emission or confused emission from many
compact sources. The evidence seems to suggest that that there may be
intense highly concentrated star formation activity in this region.

\subsection{Foreground ISM Free-free Opacities}
\label{se:ISMFF}

The results of the spectral fitting in Section~\ref{se:specanal} give
overall estimated free--free opacities for each source which are the
sum of a foreground ISM opacity and a local opacity due to an ionized
CSM. Despite this they can be used to constrain properties of the
ionized ISM because the fitted $\tau_{18}$ gives a firm {\it upper
  limit} on the ionized ISM opacity at 18~cm $\tau_{\rm ISM, 18}$
along each line of sight. Furthermore, it was argued in
Section~\ref{se:SNe} that sources of classes R, S and A are probably
RSNe with significant CSM FFA while class L sources were most likely
SNRs with no CSM absorption. If we accept this conclusions then the
measured opacities for class L sources are not just upper limits to
$\tau_{\rm ISM, 18}$ but actual measurements. Amongst the six type L
sources with good fits four were found to have one sigma upper limits
on opacity of $0.3$.  For these lines of sight (taking into account
also possible systematic calibration errors) we estimate that
$\tau_{\rm ISM,18}<0.6$ (see Figure~\ref{fi:taualpha}). The other two
sources had $\tau_{18} = 1.2$ and $1.5$ respectively.  Although we
only have small number statistics these results are suggestive that in
the western nucleus of Arp~220 there is a patchy FFA ISM with a median
opacity of less than one. It is presently impossible to give any
estimate for the eastern nucleus, because of the lack of good spectral
fits to any L type source.

It is interesting to compare our results with those of \citet{ANAN00}
who modelled the ionized thermal gas component in Arp~220 by jointly
fitting to radio recombination lines and the continuum spectrum. The
predictions of this model have recently been confirmed by VLA
observations of another recombination line \citep{RODRIC05}.
\citet{ANAN00} present an integrated continuum spectrum (encomposing
both nuclei but dominated at most frequencies by the western nucleus).
This spectrum shows a steep spectrum non thermal power law from 3 to
30~GHz. From 300~MHz to 2~GHz there is flat spectrum region and below
300~MHz a sharp cutoff. \citet{ANAN00} invoke a non thermal
synchrotron continuum plus three ionized gas components (A1, A2 and D)
to jointly explain this spectrum and observed radio recombination
lines. The A2 component (from high pressure HII regions) has a very
low covering factor and in continuum effects only high frequencies.
The A1 and D components have $\tau_{\rm ISM,18}=0.97$ and $0.02$ with
area covering factors of $0.7$ and $1$ respectively. These two
components explain the flat region below 2~GHz (where A1 starts to
become optically thick) and the sharp drop below 300~MHz (where D
becomes optically thick). The estimated opacities derived for our
compact L source seem broadly to be consistent with this model, with
some sources having little foreground FFA and others a moderate
amount.

\subsection{Star-formation Rates and the Nature of Arp~220's
  Starburst}
\label{se:SFR}

As noted by \citet{LONSDALE06} the SFR inferred from the FIR
luminosity is consistent with the rate of appearance of new radio
sources ($4\pm2$~\per{year}) {\em if} all or most core collapse SNe
give rise to a detectable radio source. However there is evidence from
their luminosity, radio spectra and evolutionary timescale confirming
that these new sources are Type~IIn SNe. This class of SN are thought
to be relatively rare in normal galaxies \citep{CAPPELLARO97}
comprising only 2\% of core collapse SN (though due to their extreme
luminosities, they are much more common in compilations of detected
SNe). Additionally Avishay Gal-Yam et al. (astro-ph/0608029) have
recently identified the progenitor of the Type~IIn SN2005gl as a
luminous blue variable (LBV) star. Such stars are known to have
initial masses in excess of \about 80~\Msun. If a standard IMF is
assumed then the fraction of core collapse SNe with progenitors in
this mass range is \about 3\%, similar to the expected fraction of
Type~II SNe, implying that all or most Type~IIn progenitors are LBVs.
If this is true then under conventional RSNe and starburst models the
observed rate of appearance of new compact radio sources could be up
to 50 times larger than the number predicted from the SFR. A full
discussion of this issue will be left to a future paper but it is
interesting to speculate on possible resolutions of this apparent
inconsistency.

For instance, the density in the nuclear region of Arp~220 could be so
high that the adopted paradigm of an early SN phase followed by a SNR
phase does not apply. Instead every core collapse SN interacts
directly with the dense ISM and gives a bright radio source. Hence the
observed rate of new radio sources matches the prediction from the SFR
\citep{LONSDALE06}. However, so far it seems that the properties of
the new sources are consistent with standard RSNe interacting with
their stellar winds. In this case a possible explanation is that the
IMF in Arp~220 has an unusual overabundance of very massive stars.
Such a scenario for Arp~220 would be very interesting since it would
suggest a difference in star formation mechanisms compared to those
found in galactic disks. Nevertheless, a recent review \citep{ELM05}
points out that there are considerable uncertainities in measuring the
IMF in massive star clusters. Therefore, there is no decisive evidence
in favour of a top heavy IMF in starburst galaxies.

Another possible scenario consists of a normal IMF but that the late
stages in the evolution of massive stars is fundamentally different in
Arp~220 due to the very high (stellar) density ULIRG environment. It
has been predicted that binary systems will give more Type~IIb/IIL
sources at the expense of radio-weak Type~IIP
\citep{NOMOTO96,CHEVALIER06P}. Possibly some mechanism might exist
such that a higher fraction of multiple star systems or close stellar
encounters also dramatically increases the fraction of the extremely
luminous Type~IIn events.

One last possibility is that the SFR of the starburst is highly
variable on the timescale of massive star lifetimes, and we are now
observing Arp~220 just when the massive stars formed in a previous
burst are exploding as SNe. Based on their observations of radio
recombination lines, one of the scenarios discussed in \citet{ANAN00}
is that the star formation in Arp~220 could be occuring in intense but
very short bursts, each lasting a few times \E{5}~year with a SFR in
the order of \E{3}~\Msun~\per{year}, separated by a few times
\E{6}~year.

To check whether this scenario can provide enough massive stars to
explain the observed luminous RSN rate, consider that the previous
burst lasted 3\x\E{5}~years with a SFR of 10 times the averaged SFR
estimated by \citet{ANAN00}, i.e. 2400~\Msun~\per{year}, thus giving a
total mass of 7\x\E{8}~\Msun\ in new stars. If a Salpeter IMF with
power law exponent of -2.3 and a low mass cutoff of 0.5~\Msun\ is
assumed \citep{KROUPA02}, then 4.5\x\E{5} of these stars are estimated
to be more massive than $80$~\Msun (note that no upper mass limit was
used; if in contrast, an upper limit of 120~\Msun\ is assumed, this
estimate is reduced by a factor of 2). If the previous burst occured
one LBV lifetime ago \citep[$3\times 10^6$~years,][]{MASSEY01} these
massive stars would be exploding over a period defined by the sum of
the burst length and the stellar lifetimes. For such massive stars,
the H-burning time is known to be a weak function of mass
\citep{MASSEY01}, therefore, assuming a dispersion in their lifetimes
of 10\% (which is comparable to the burst duration) then all these
massive stars would explode over a period of 6\x\E{5}~years implying a
luminous RSN rate of 0.75~\per{year}; which is of the order of the
observed rate.

The above calculation shows that despite the many uncertainties the
short burst explanation for the large rate of powerful RSNe is
feasible. However a possible problem is that it requires that
fortuitously there was a burst exactly 3\x\E{6}~ years ago to provide
the RSNe we see plus another ongoing burst to provide the large number
of very young ($<$\E{5}~years) overpressured HII regions required by
\citet{ANAN00}; moreover this would have to be true for both nuclei.
\citet{SCOVILLE97} argue that the two nuclei are embedded within a
larger scale disk and are mutually orbiting each other. We can
speculate that one way to obtain the needed star formation history
would be if their orbit were at least somwhat elliptical and bursts
were triggered by their periodic periapsis passage. Based on the
seperation of the nuclei and their relative systemic velocities
\citet{SCOVILLE97} argue for an orbital period for the double nucleus
of \about4.6\x\E{6}~years. Although this period estimate is uncertain
encouragingly it is comparable to LBV lifetimes.

\section{CONCLUSIONS AND FUTURE PROSPECTS}\label{se:conclusion}

The main conclusions of this paper are as follows:

\paragraph{1} For the first time we have detected the compact radio
sources in Arp~220 at wavelengths shorter than 18~cm (see
Section~\ref{se:VLBA}). Previous failures to detect the compact
sources at 6~cm appear to be largely due to using a phase calibrator
which was too distant, possibly combined with the observations being
made near the maximum of the solar cycle. We also uncovered an
astrometric error of \about100~mas in declination in previously
published absolute OH maser and 18~cm continuum VLBI positions (see
Section ~\ref{se:astrometry}).

\paragraph{2} A total of 18 sources (all but three in the western
nucleus) are clearly detected at short wavelengths (see
Table~\ref{ta:source_fluxes}). For these we construct estimated radio
spectra from our simultaneous measurements at 3.6, 6, 13~cm and a
non-simultaneous 18~cm measurement. A wide variety of spectral shapes
are seen ranging from inverted through peaked to steep (see
Section~\ref{se:spectra}). Most can be well fitted by a simple model
consisting of a synchrotron power law plus foreground free-free
absorption (see Section~\ref{se:astrometry}).

\paragraph{3} There is evidence that younger sources have higher
free-free absorption consistent with what is expected from RSN models
(see Section~\ref{se:SNe}). In particular four sources which have
recently appeared at 18~cm seem well fitted by such models. Newly
cataloged sources seen only at short wavelength may be even younger SN
or old SNR with especially large foreground ISM opacity. In general
for sources which were not detected in the original discovery
observations of \citet{SMITH98} the combination of their fitted
free--free opacities, luminosities and age limits are consistent with
what is expected for Type~IIn RSNe

\paragraph{4} The bright 18~cm sources originally detected by
\citet{SMITH98} in addition to showing little time variability also
have lower fitted free--free opacities than the other sources. The
underlying synchrotron spectral index of these sources shows a wide
range and contains both relatively flat $\alpha>-0.3$ and steep
spectrum ($\alpha<-1$) sources (see Section~\ref{se:probnat}).  It
seems difficult to interpret these sources as standard RSNe because of
their combined age and luminosity (see Section~\ref{se:SNe}).  Models
in which they are instead young SNR interacting with the surrounding
ISM are more successful. If they follow the same luminosity-size
relation as the SNRs in M82 their luminosities imply diameters of
$0.2$~pc and surrounding ISM of densities in the range $\E{4}<n_{\rm
  ISM}<\E{5}$~cm\tothe{-3}. Ages for these sources would be only
30--50~years, similar to those estimated by \citet{LONSDALE06}.

\paragraph{5} One source (W42) is probably resolved at 3.6~cm
(diameter 0.86~pc). Estimates of the minimum total energy in particles
and fields (see Section~\ref{se:W42}) suggest that the initial kinetic
energy of the supernovae was larger than in typical core collapse
supernova and could be as high as \E{52}~ergs.

\paragraph{6} Source W33 shows a complex spectrum and has evidence for
significant epoch to epoch variability (see Section~\ref{se:W33}). It
is located in a crowded region close to a patch of apparently high
brightness extended 18~cm emission. There is possible evidence of
extended structure at 13~cm but this may be caused by imaging
artifacts.  The area around W33 has a high density of compact sources
and is a highly active star formation region.

\paragraph{7} Based on the fitted free-free absorption toward the
stable SNR candidates in the western nucleus we are able constrain the
opacity due to the ionized component of the ISM. Four sources have
opacities at 18~cm of $\tau_{\rm 18,ISM} < 0.6$, the other two have
$\tau_{\rm 18,ISM} = 1.2$ and $1.5$ respectively. These results are in
good agreement with the model of \citet{ANAN00} which postulates two
phases of ionized gas affecting absorption at low frequency, one with
covering factor $0.7$ and opacity $\tau_{\rm 18,ISM}= 1$ the other
with a covering factor of 1 and $\tau_{\rm 18,ISM}= 0.02$.

\paragraph{8} \citet{LONSDALE06} found that the rate of appearance of
new radio sources approximately equals that expected from the FIR
derived SFR if every core-collapse SN gives rise to a bright radio
source.  However slowly evolving SN sources as bright as those
detected in Arp~220 belong to the Type~IIn class which are thought to
be relatively rare in galactic disks. Consistent with this is recent
identification of Type~IIn progenitors as extremely massive LBV stars
($>$80\Msun). In Section~\ref{se:SFR} we discuss possible explanations
to this apparent conflict. Amongst the possibilities are that a
totally different RSN paradigm applies in ULIRGs, that a top heavy
stellar IMF or non-standard stellar evolution applies. A final
possibility is that the recent starburst activity occurred in a very
short but intense burst which we are now observing just as the most
massive stars explode as SNe.
   
\paragraph{9} In Section \ref{se:flat} we noted the outside
possibility that one of the flatter spectrum source could be an AGN,
likewise the source with unusual spectrum, W33 (see Section
\ref{se:W33}) might be another possible candidate. However based on
present data there is no convincing case of any source which requires
such an AGN explanation.
  
\vskip 1cm

The detection of the Arp~220 compact sources at short wavelengths
opens up many interesting future observational possibilities. The
observations presented here only scratch the surface. Recently
conducted (May 2006) and proposed global VLBI observations using the
world's largest telescopes and largest bandwidths will give images 10
times more sensitive than those presented in this paper. These data
will help determine the spectral properties of the weaker sources and
constrain systematic differences with luminosity. With a larger number
of detected sources we can also test for possible differences in the
spatial distribution of sources with different spectra. Observations
at the shortest wavelengths can be used to see if our apparently flat
spectrum remain so to the highest frequencies. Additionally, because
SNe evolve much faster at short wavelengths, source monitoring will
produce much faster results than at 18~cm. Multi-wavelength light
curve monitoring will allow us to make detailed SN fits for stellar
mass loss properties (indirectly probing stellar evolution).
Additionally such fitting will give more accurate estimates of
foreground FFA allowing us to map the spatial distribution of the
ionized ISM.

Finally future short wavelength observations may be able to resolve or
set useful limits on source sizes. One candidate SNR source (W42) has
probably already been resolved. A SN in free expansion at \E{4}~\kms\
will after 5~years have diameter of 0.1~pc (0.3~mas), which is within
the current observational resolving capabilities. Likewise we can
check the prediction that those sources we have identified as SNR have
diameters of 0.2~pc (0.6~mas). Finally if such SNR follow the expected
Luminosity$-D$ correlation then the weaker sources are expected to be
larger, a prediction which can be checked by future more sensitive
observations.

\acknowledgements{RP acknowledges a Chalmers University PhD student
  stipend and JC a Swedish VR grant. This work was partially supported
  by NSF grant AST-0352953 to Haystack Observatory. The European VLBI
  Network is a joint facility of European, Chinese, South African and
  other radio astronomy institutes funded by their national research
  councils. The Arecibo Observatory is the principal facility of the
  National Astronomy and Ionosphere Center, which is operated by the
  Cornell University under a cooperative agreement with the National
  Science Foundation. The National Radio Astronomy Observatory is a
  facility of the National Science Foundation operated under
  cooperative agreement by Associated Universities, Inc. We thank the
  anonymous referee for his/her valuable comments and suggestions.}

\bibliographystyle{apj}
\bibliography{ms}

\begin{thebibliography}{61}
\expandafter\ifx\csname natexlab\endcsname\relax\def\natexlab#1{#1}\fi

\bibitem[{{Alberdi} {et~al.}(2006){Alberdi}, {Colina}, {Torrelles}, {Panagia},
  {Wilson}, \& {Garrington}}]{ALBERDI06}
{Alberdi}, A., {Colina}, L., {Torrelles}, J.~M., {Panagia}, N., {Wilson},
  A.~S., \& {Garrington}, S.~T. 2006, \apj, 638, 938

\bibitem[{{Allen} \& {Kronberg}(1998)}]{ALLEN98}
{Allen}, M.~L. \& {Kronberg}, P.~P. 1998, \apj, 502, 218

\bibitem[{{Anantharamaiah} {et~al.}(2000){Anantharamaiah}, {Viallefond},
  {Mohan}, {Goss}, \& {Zhao}}]{ANAN00}
{Anantharamaiah}, K.~R., {Viallefond}, F., {Mohan}, N.~R., {Goss}, W.~M., \&
  {Zhao}, J.~H. 2000, \apj, 537, 613

\bibitem[{{Asvarov}(2006)}]{ASVAROV06}
{Asvarov}, A.~I. 2006, \aap, 459, 519

\bibitem[{{Bartel} {et~al.}(2002){Bartel}, {Bietenholz}, {Rupen}, {Beasley},
  {Graham}, {Altunin}, {Venturi}, {Umana}, {Cannon}, \& {Conway}}]{BARTEL02}
{Bartel}, N., {Bietenholz}, M.~F., {Rupen}, M.~P., {Beasley}, A.~J., {Graham},
  D.~A., {Altunin}, V.~I., {Venturi}, T., {Umana}, G., {Cannon}, W.~H., \&
  {Conway}, J.~E. 2002, \apj, 581, 404

\bibitem[{{Bartel} {et~al.}(1987){Bartel}, {Ratner}, {Rogers}, {Shapiro},
  {Bonometti}, {Cohen}, {Gorenstein}, {Marcaide}, \& {Preston}}]{BARTEL87}
{Bartel}, N., {Ratner}, M.~I., {Rogers}, A.~E.~E., {Shapiro}, I.~I.,
  {Bonometti}, R.~J., {Cohen}, N.~L., {Gorenstein}, M.~V., {Marcaide}, J.~M.,
  \& {Preston}, R.~A. 1987, \apj, 323, 505

\bibitem[{{Beasley} \& {Conway}(1995)}]{BEASLEY95}
{Beasley}, A.~J. \& {Conway}, J.~E. 1995, in ASP Conf. Ser. 82: Very Long
  Baseline Interferometry and the VLBA, ed. J.~A. {Zensus}, P.~J. {Diamond}, \&
  P.~J. {Napier}, 328--+

\bibitem[{{Beck} \& {Krause}(2005)}]{BECK05}
{Beck}, R. \& {Krause}, M. 2005, Astronomische Nachrichten, 326, 414

\bibitem[{{Berezhko} \& {V{\"o}lk}(2004)}]{BEREZHKO04}
{Berezhko}, E.~G. \& {V{\"o}lk}, H.~J. 2004, \aap, 427, 525

\bibitem[{{Beswick} {et~al.}(2006){Beswick}, {Riley}, {Marti-Vidal}, {Pedlar},
  {Muxlow}, {McDonald}, {Wills}, {Fenech}, \& {Argo}}]{BESWICK06}
{Beswick}, R.~J., {Riley}, J.~D., {Marti-Vidal}, I., {Pedlar}, A., {Muxlow},
  T.~W.~B., {McDonald}, A.~R., {Wills}, K.~A., {Fenech}, D., \& {Argo}, M.~K.
  2006, \mnras, 369, 1221

\bibitem[{{Bietenholz} {et~al.}(2002){Bietenholz}, {Bartel}, \&
  {Rupen}}]{BIET02}
{Bietenholz}, M.~F., {Bartel}, N., \& {Rupen}, M.~P. 2002, \apj, 581, 1132

\bibitem[{{Bietenholz} {et~al.}(2004){Bietenholz}, {Bartel}, \&
  {Rupen}}]{BIET04}
---. 2004, Science, 304, 1947

\bibitem[{{Cappellaro} {et~al.}(1997){Cappellaro}, {Turatto}, {Tsvetkov},
  {Bartunov}, {Pollas}, {Evans}, \& {Hamuy}}]{CAPPELLARO97}
{Cappellaro}, E., {Turatto}, M., {Tsvetkov}, D.~Y., {Bartunov}, O.~S.,
  {Pollas}, C., {Evans}, R., \& {Hamuy}, M. 1997, \aap, 322, 431

\bibitem[{{Chevalier}(2006)}]{CHEVALIER06P}
{Chevalier}, R. 2006, in Proceedings of 2006 STScI May Symposium on Massive
  Stars

\bibitem[{{Chevalier}(1982{\natexlab{a}})}]{CHEVALIER82b}
{Chevalier}, R.~A. 1982{\natexlab{a}}, \apjl, 259, L85

\bibitem[{{Chevalier}(1982{\natexlab{b}})}]{CHEVALIER82a}
---. 1982{\natexlab{b}}, \apj, 259, 302

\bibitem[{{Chevalier}(1998)}]{CHEVALIER98}
---. 1998, \apj, 499, 810

\bibitem[{{Chevalier}(2005)}]{CHEVALIER05}
---. 2005, \apj, 619, 839

\bibitem[{{Chevalier} \& {Fransson}(1994)}]{CHEVALIER94}
{Chevalier}, R.~A. \& {Fransson}, C. 1994, \apj, 420, 268

\bibitem[{{Chevalier} \& {Fransson}(2001)}]{CHEVALIER01}
---. 2001, \apjl, 558, L27

\bibitem[{{Chevalier} {et~al.}(2006){Chevalier}, {Fransson}, \&
  {Nymark}}]{CHEVALIER06}
{Chevalier}, R.~A., {Fransson}, C., \& {Nymark}, T.~K. 2006, \apj, 641, 1029

\bibitem[{{Chevalier} {et~al.}(2004){Chevalier}, {Li}, \&
  {Fransson}}]{CHEVALIER04}
{Chevalier}, R.~A., {Li}, Z.-Y., \& {Fransson}, C. 2004, \apj, 606, 369

\bibitem[{{Corbett} {et~al.}(2003){Corbett}, {Kewley}, {Appleton},
  {Charmandaris}, {Dopita}, {Heisler}, {Norris}, {Zezas}, \&
  {Marston}}]{COLAS2}
{Corbett}, E.~A., {Kewley}, L., {Appleton}, P.~N., {Charmandaris}, V.,
  {Dopita}, M.~A., {Heisler}, C.~A., {Norris}, R.~P., {Zezas}, A., \&
  {Marston}, A. 2003, \apj, 583, 670

\bibitem[{{Corbett} {et~al.}(2002){Corbett}, {Norris}, {Heisler}, {Dopita},
  {Appleton}, {Struck}, {Murphy}, {Marston}, {Charmandaris}, {Kewley}, \&
  {Zezas}}]{COLAS1}
{Corbett}, E.~A., {Norris}, R.~P., {Heisler}, C.~A., {Dopita}, M.~A.,
  {Appleton}, P., {Struck}, C., {Murphy}, T., {Marston}, A., {Charmandaris},
  V., {Kewley}, L., \& {Zezas}, A.~L. 2002, \apj, 564, 650

\bibitem[{{Cowsik} \& {Sarkar}(1984)}]{COWSIK84}
{Cowsik}, R. \& {Sarkar}, S. 1984, \mnras, 207, 745

\bibitem[{{Dopita} {et~al.}(2005){Dopita}, {Groves}, {Fischera}, {Sutherland},
  {Tuffs}, {Popescu}, {Kewley}, {Reuland}, \& {Leitherer}}]{DOPITA05}
{Dopita}, M.~A., {Groves}, B.~A., {Fischera}, J., {Sutherland}, R.~S., {Tuffs},
  R.~J., {Popescu}, C.~C., {Kewley}, L.~J., {Reuland}, M., \& {Leitherer}, C.
  2005, \apj, 619, 755

\bibitem[{{Draine} \& {Woods}(1991)}]{DRAINE91}
{Draine}, B.~T. \& {Woods}, D.~T. 1991, \apj, 383, 621

\bibitem[{{Elmegreen}(2005)}]{ELM05}
{Elmegreen}, B.~G. 2005, in ASSL Vol. 329: Starbursts: From 30 Doradus to Lyman
  Break Galaxies, ed. R.~{de Grijs} \& R.~M. {Gonz{\'a}lez Delgado}, 57--+

\bibitem[{{Huang} {et~al.}(1994){Huang}, {Thuan}, {Chevalier}, {Condon}, \&
  {Yin}}]{HUANG94}
{Huang}, Z.~P., {Thuan}, T.~X., {Chevalier}, R.~A., {Condon}, J.~J., \& {Yin},
  Q.~F. 1994, \apj, 424, 114

\bibitem[{{Kronberg} {et~al.}(1985){Kronberg}, {Biermann}, \&
  {Schwab}}]{KRON85}
{Kronberg}, P.~P., {Biermann}, P., \& {Schwab}, F.~R. 1985, \apj, 291, 693

\bibitem[{{Kroupa}(2002)}]{KROUPA02}
{Kroupa}, P. 2002, Science, 295, 82

\bibitem[{{Longair}(1994)}]{LONGAIRV2}
{Longair}, M.~S. 1994, {High energy astrophysics. Vol.2: Stars, the galaxy and
  the interstellar medium} (Cambridge: Cambridge University Press, |c1994, 2nd
  ed.)

\bibitem[{{Lonsdale} {et~al.}(2006){Lonsdale}, {Diamond}, {Thrall}, {Smith}, \&
  {Lonsdale}}]{LONSDALE06}
{Lonsdale}, C.~J., {Diamond}, P.~J., {Thrall}, H., {Smith}, H.~E., \&
  {Lonsdale}, C.~J. 2006, \apj, 647, 185

\bibitem[{{Massey} {et~al.}(2001){Massey}, {DeGioia-Eastwood}, \&
  {Waterhouse}}]{MASSEY01}
{Massey}, P., {DeGioia-Eastwood}, K., \& {Waterhouse}, E. 2001, \aj, 121, 1050

\bibitem[{{McDonald} {et~al.}(2001){McDonald}, {Muxlow}, {Pedlar}, {Garrett},
  {Wills}, {Garrington}, {Diamond}, \& {Wilkinson}}]{McDONALD01}
{McDonald}, A.~R., {Muxlow}, T.~W.~B., {Pedlar}, A., {Garrett}, M.~A., {Wills},
  K.~A., {Garrington}, S.~T., {Diamond}, P.~J., \& {Wilkinson}, P.~N. 2001,
  \mnras, 322, 100

\bibitem[{{McDonald} {et~al.}(2002){McDonald}, {Muxlow}, {Wills}, {Pedlar}, \&
  {Beswick}}]{McDONALD02}
{McDonald}, A.~R., {Muxlow}, T.~W.~B., {Wills}, K.~A., {Pedlar}, A., \&
  {Beswick}, R.~J. 2002, \mnras, 334, 912

\bibitem[{{Muxlow} {et~al.}(1994){Muxlow}, {Pedlar}, {Wilkinson}, {Axon},
  {Sanders}, \& {de Bruyn}}]{MUX94}
{Muxlow}, T.~W.~B., {Pedlar}, A., {Wilkinson}, P.~N., {Axon}, D.~J., {Sanders},
  E.~M., \& {de Bruyn}, A.~G. 1994, \mnras, 266, 455

\bibitem[{{Narayanan} {et~al.}(2005){Narayanan}, {Groppi}, {Kulesa}, \&
  {Walker}}]{NARAYANAN05}
{Narayanan}, D., {Groppi}, C.~E., {Kulesa}, C.~A., \& {Walker}, C.~K. 2005,
  \apj, 630, 269

\bibitem[{{Natta} \& {Panagia}(1984)}]{NATTA84}
{Natta}, A. \& {Panagia}, N. 1984, \apj, 287, 228

\bibitem[{{Neff} {et~al.}(2004){Neff}, {Ulvestad}, \& {Teng}}]{NEFF04}
{Neff}, S.~G., {Ulvestad}, J.~S., \& {Teng}, S.~H. 2004, \apj, 611, 186

\bibitem[{{Nomoto} {et~al.}(1996){Nomoto}, {Iwamoto}, {Suzuki}, {Pols},
  {Yamaoka}, {Hashimoto}, {Hoflich}, \& {van den Heuvel}}]{NOMOTO96}
{Nomoto}, K., {Iwamoto}, K., {Suzuki}, T., {Pols}, O.~R., {Yamaoka}, H.,
  {Hashimoto}, M., {Hoflich}, P., \& {van den Heuvel}, E.~P.~J. 1996, in IAU
  Symp. 165: Compact Stars in Binaries, ed. J.~{van Paradijs}, E.~P.~J. {van
  den Heuvel}, \& E.~{Kuulkers}, 119--+

\bibitem[{{Ostrowski}(1999)}]{OSTROWSKI99}
{Ostrowski}, M. 1999, \aap, 345, 256

\bibitem[{{Parra} {et~al.}(2005){Parra}, {Conway}, {Appleton}, \&
  {Pihlstr{\"o}m}}]{BADHONNEF}
{Parra}, R., {Conway}, J., {Appleton}, P., \& {Pihlstr{\"o}m}, Y. 2005, in AIP
  Conf. Proc. 783, 241--244

\bibitem[{{Petrov} {et~al.}(2005){Petrov}, {Kovalev}, {Fomalont}, \&
  {Gordon}}]{PETROV05}
{Petrov}, L., {Kovalev}, Y.~Y., {Fomalont}, E., \& {Gordon}, D. 2005, \aj, 129,
  1163

\bibitem[{{Pradel} {et~al.}(2005){Pradel}, {Charlot}, \& {Lestrade}}]{PRADEL05}
{Pradel}, N., {Charlot}, P., \& {Lestrade}, J.-F. 2005, in ASP Conf. Ser. 340:
  Future Directions in High Resolution Astronomy, ed. J.~{Romney} \& M.~{Reid},
  538--+

\bibitem[{{Rodr{\'{\i}}guez-Rico} {et~al.}(2005){Rodr{\'{\i}}guez-Rico},
  {Goss}, {Viallefond}, {Zhao}, {G{\'o}mez}, \& {Anantharamaiah}}]{RODRIC05}
{Rodr{\'{\i}}guez-Rico}, C.~A., {Goss}, W.~M., {Viallefond}, F., {Zhao}, J.-H.,
  {G{\'o}mez}, Y., \& {Anantharamaiah}, K.~R. 2005, \apj, 633, 198

\bibitem[{{Rovilos} {et~al.}(2003){Rovilos}, {Diamond}, {Lonsdale}, {Lonsdale},
  \& {Smith}}]{ROVILOS03}
{Rovilos}, E., {Diamond}, P.~J., {Lonsdale}, C.~J., {Lonsdale}, C.~J., \&
  {Smith}, H.~E. 2003, \mnras, 342, 373

\bibitem[{{Rovilos} {et~al.}(2005){Rovilos}, {Diamond}, {Lonsdale}, {Smith}, \&
  {Lonsdale}}]{ROVILOS05}
{Rovilos}, E., {Diamond}, P.~J., {Lonsdale}, C.~J., {Smith}, H.~E., \&
  {Lonsdale}, C.~J. 2005, \mnras, 359, 827

\bibitem[{{Scoville} {et~al.}(1997){Scoville}, {Yun}, \& {Bryant}}]{SCOVILLE97}
{Scoville}, N.~Z., {Yun}, M.~S., \& {Bryant}, P.~M. 1997, \apj, 484, 702

\bibitem[{{Smith} {et~al.}(1998){Smith}, {Lonsdale}, {Lonsdale}, \&
  {Diamond}}]{SMITH98}
{Smith}, H.~E., {Lonsdale}, C.~J., {Lonsdale}, C.~J., \& {Diamond}, P.~J. 1998,
  \apjl, 493, L17+

\bibitem[{{Soderberg} {et~al.}(2005){Soderberg}, {Kulkarni}, {Berger},
  {Chevalier}, {Frail}, {Fox}, \& {Walker}}]{SODERBERG05}
{Soderberg}, A.~M., {Kulkarni}, S.~R., {Berger}, E., {Chevalier}, R.~A.,
  {Frail}, D.~A., {Fox}, D.~B., \& {Walker}, R.~C. 2005, \apj, 621, 908

\bibitem[{{Thompson} {et~al.}(2001){Thompson}, {Moran}, \&
  {Swenson}}]{INT_BOOK}
{Thompson}, A.~R., {Moran}, J.~M., \& {Swenson}, Jr., G.~W. 2001,
  {Interferometry and Synthesis in Radio Astronomy, 2nd Edition} (A
  Wiley-Interscience publication. ISBN : 0471254924)

\bibitem[{{Ulvestad} \& {Antonucci}(1993)}]{UA93}
{Ulvestad}, J.~S. \& {Antonucci}, R.~R.~J. 1993, in Bulletin of the American
  Astronomical Society, 842--+

\bibitem[{{Uro{\v s}evi{\'c}} {et~al.}(2005){Uro{\v s}evi{\'c}}, {Pannuti},
  {Duric}, \& {Theodorou}}]{UROSEVIC05}
{Uro{\v s}evi{\'c}}, D., {Pannuti}, T.~G., {Duric}, N., \& {Theodorou}, A.
  2005, \aap, 435, 437

\bibitem[{{van Marle} {et~al.}(2004){van Marle}, {Langer}, \&
  {Garc{\'{\i}}a-Segura}}]{VANMARLE04}
{van Marle}, A.~J., {Langer}, N., \& {Garc{\'{\i}}a-Segura}, G. 2004, in
  Revista Mexicana de Astronomia y Astrofisica Conference Series, ed.
  G.~{Garcia-Segura}, G.~{Tenorio-Tagle}, J.~{Franco}, \& H.~W. {Yorke},
  136--139

\bibitem[{{Weiler} {et~al.}(2002){Weiler}, {Panagia}, {Montes}, \&
  {Sramek}}]{WEILER_REVIEW}
{Weiler}, K.~W., {Panagia}, N., {Montes}, M.~J., \& {Sramek}, R.~A. 2002,
  \araa, 40, 387

\bibitem[{{Wheeler} {et~al.}(1980){Wheeler}, {Mazurek}, \&
  {Sivaramakrishnan}}]{WHEELER80}
{Wheeler}, J.~C., {Mazurek}, T.~J., \& {Sivaramakrishnan}, A. 1980, \apj, 237,
  781

\bibitem[{{Wills} {et~al.}(1997){Wills}, {Pedlar}, {Muxlow}, \&
  {Wilkinson}}]{WILLS97}
{Wills}, K.~A., {Pedlar}, A., {Muxlow}, T.~W.~B., \& {Wilkinson}, P.~N. 1997,
  \mnras, 291, 517

\bibitem[{{Woosley} \& {Bloom}(2006)}]{WOOSLEY06}
{Woosley}, S.~E. \& {Bloom}, J.~S. 2006, \araa, 44, 507

\bibitem[{{Wrobel} \& {Ho}(2006)}]{WROBEL06}
{Wrobel}, J.~M. \& {Ho}, L.~C. 2006, \apjl, 646, L95

\bibitem[{{Wrobel} {et~al.}(2000){Wrobel}, {Walker}, \& {Benson}}]{MEMO24}
{Wrobel}, J.~M., {Walker}, R.~C., \& {Benson}, J.~M. 2000, VLBA correlator memo
  No 24

\end{thebibliography}

%%%%%%%%%%%%%%%%%%%%%%%%%%%%%%%%%%%%%%%%%%%%%%%%%%%%%%%%%%%%%%%%%%%%%%
\begin{deluxetable}{rlrccc}
\tabletypesize{\scriptsize}
\tablecolumns{6}
\tablewidth{1\hsize}
\tablecaption{VLBI observations of Arp~220}
\tablehead{
\colhead{Epoch}&
\colhead{Code}&
\colhead{$\lambda$}&
\colhead{Array}&
\colhead{$\sigma_{\lambda}$}&
\colhead{Beam Size}
\\
&
&
\colhead{(cm)}&
&
\colhead{(\uJy~\per{beam})}&
\colhead{(mas$^{2}$)}}
\startdata
2003.01 &BN022  & 6.02  & VLBA     &147.01& 4.4\x 2.0\\
2003.85 &GD17A\tablenotemark{a}  &18.18  & GVLBI     & \phantom{14}9.00 & 5.9\x 2.7\\
2005.16&EP049  & 6.02  & Eb, Wb, Ar & \phantom{1}42.08 & \phantom{\tablenotemark{c}}200\x200\tablenotemark{c}\\
2005.18&GD17B\tablenotemark{b}  &18.18  & GVLBI     & \phantom{14}9.00 & 7.6\x 2.9\\
2006.02&BP129  &13.26  & VLBA     &129.54 & 6.6\x 3.6\\
           &       & 6.02  &          & \phantom{1}86.02 & 3.3\x 1.8\\
           &       & 3.56  &          & \phantom{1}86.73 & 3.1\x 1.7\\
\enddata
\tablenotetext{a}{\citet{LONSDALE06}}
\tablenotetext{b}{Thrall et al., in preparation}
\tablenotetext{c}{Sensitive only to $<$2~mas features. See text for details.}
\label{ta:epochs}
\end{deluxetable}

%%%%%%%%%%%%%%%%%%%%%%%%%%%%%%%%%%%%%%%%%%%%%%%%%%%%%%%%%%%%%%%%%%%%%%

%%%%%%%%%%%%%%%%%%%%%%%%%%%%%%%%%%%%%%%%%%%%%%%%%%%%%%%%%%%%%%%%%%%%%%
\begin{deluxetable}{cccrrrrrrrrrc}
\tabletypesize{\scriptsize}
\tablecolumns{13}
\tablewidth{0pt}
\tablecaption{Properties of detected radio sources and measured fluxes in \uJy}
\tablehead{
\colhead{}&
\colhead{}&
\colhead{$\alpha_{2000}$}&
\colhead{$\delta_{2000}$}&
\colhead{GD17A}&
\colhead{GD17B}&
\colhead{}&
\colhead{BP129}&
\colhead{}&
\colhead{}&
\colhead{}&
\colhead{}&
\colhead{}
\\
\cline{7-9}
\colhead{Name}&
\colhead{SN}&
\colhead{15\hr34\min...}&
\colhead{23\deg30\arcmin...}&
\colhead{18 cm}&
\colhead{18 cm}&
\colhead{13 cm}&
\colhead{6 cm}&
\colhead{3.6 cm}&
\colhead{$\alpha$}&
\colhead{$\tau_{18}$}&
\colhead{$S_{\rm sy}$}&
\colhead{Class}
\\
\colhead{(1)}&
\colhead{(2)}&
\colhead{(3)}&
\colhead{(4)}&
\colhead{(5)}&
\colhead{(6)}&
\colhead{(7)}&
\colhead{(8)}&
\colhead{(9)}&
\colhead{(10)}&
\colhead{(11)}&
\colhead{(12)}&
\colhead{(13)}
}
\startdata
W8  &3    &57\rlap.\sec2361&11\rlap.\arcsec4318 & 849.2 &828.2     &483.8    &288.0  &16.2   &$-$1.27 & 0.00 &  809.7  & L \\
W10 &4    &57\rlap.\sec2307&11\rlap.\arcsec5022 & 408.8 &463.1     &417.3    &311.9  &274.4  &$-$0.29 & 0.00 &  440.9  & L \\
W11 &     &57\rlap.\sec2230&11\rlap.\arcsec5015 & 175.4 &379.3     &572.6    &382.4  &357.5  &$-$0.72 & 0.68 &  989.7  & R \\
W12 &     &57\rlap.\sec2295&11\rlap.\arcsec5244 & 645.6 &940.9     &1011.4   &458.1  &341.0  &$-$0.72 & 0.00 &  989.8  & R \\
W15 &     &57\rlap.\sec2253&11\rlap.\arcsec4832 & 189.3 &198.6     &579.0    &733.9  &706.6  &$-$0.72 & 2.36 & 2204.3  & A \\
W17 &6    &57\rlap.\sec2241&11\rlap.\arcsec5193 & 717.4 &693.0     &477.2    &520.8  &483.9  &$-$0.23 & 0.00 &  410.3  & L \\
W18 &7    &57\rlap.\sec2240&11\rlap.\arcsec5458 & 255.0 &261.9     &558.8    &489.5  &450.6  &$-$0.62 & 1.55 & 1236.2  & L \\
W25 &     &57\rlap.\sec2222&11\rlap.\arcsec5005 & 98.0  &499.9     &1069.1   &944.0  &648.4  &$-$0.72 & 1.03 & 2249.3  & R \\
W30 &10   &57\rlap.\sec2214&11\rlap.\arcsec4025 & 1228.0&1172.3    &127.9    &305.4  &$-$1.9 &        &      &         & L \\
W33 &11   &57\rlap.\sec2200&11\rlap.\arcsec4910 & 428.0 &338.3     &582.2    &259.3  &396.9  &        &      &         & L \\
W34 &     &57\rlap.\sec2195&11\rlap.\arcsec4919 & 160.8 &341.6     &698.8    &813.2  &742.8  &$-$0.72 & 2.08 & 2339.2  & R \\
W39 &12   &57\rlap.\sec2171&11\rlap.\arcsec4845 & 870.2 &762.2     &748.9    &279.3  &119.5  &$-$1.87 & 1.20 & 2475.1  & L \\
W42 &13   &57\rlap.\sec2123&11\rlap.\arcsec4820 & 713.5 &833.9     &743.1    &596.0  &{\it 573.7}  &$-$0.24 & 0.00 &  668.1  & L \\
W55&     &57\rlap.\sec2227&11\rlap.\arcsec4816  & \ULL	&\ULL	   &124.0    &863.9  &1147.3 &$-$0.72 & 9.72 & 5037.9  & S \\
W56&     &57\rlap.\sec2205&11\rlap.\arcsec4910  & \ULL	&\ULL	   &$-$42.3  &799.0  &749.6  &$-$0.72 & 6.78 & 3194.1  & S \\
E10 &     &57\rlap.\sec2915&11\rlap.\arcsec3350 & 80.0  &\ULL      &227.4    &692.4  &988.3  &        &      &         & A \\
E14 &     &57\rlap.\sec2868&11\rlap.\arcsec2970 & 62.1  &62.0	   &$-$100.4 &498.1  &549.2  &$-$0.72 & 8.60 & 2428.7  & A \\
E24&     &57\rlap.\sec2928&11\rlap.\arcsec3728  & \ULL	&\ULL      &114.2    &477.1  &374.4  &$-$0.72 & 4.46 & 1529.6  & S \\
\enddata
\tablecomments{{\sc Columns} --- (1): Source names from \citet{LONSDALE06} for all sources except W55, W56 and E24 which are newly detected sources and have been named using the next available "W" and "E" numbers.
(2): Names used in \citet{SMITH98} and \citet{ROVILOS05}.
(3) and (4): J2000 Right Ascencion and Declination obtained by fitting a gaussian to the highest frequency detection in the BP129 observations.
(5) and (6): 18~cm fluxes from \citet{LONSDALE06} and Thrall et al. (in  preparation) respectively. Upper limits for newly detected sources are indicated at $3\sigma$.
(7), (8) and (9): 13, 6 and 3.6~cm fluxes from the observations of this paper.
The 3.6~cm flux for source W42 (shown in italics) is given in \uJy~\per{beam} because this source is resolved.
(10): Fitted optically thin synchrotron spectral index for sources of class L (column 13) and $-0.72$ for sources in all other classes (see Section~\ref{se:fits} for details).
Sources W30, W33 and E10 resulted in poorly constrained fits so no results are given.
(11): Fitted optical depth at 18~cm. (12): Fitted optically thin synchrotron flux at 18~cm.
(13): Source variability class defined in Section~\ref{se:variability}}
\label{ta:source_fluxes}
\end{deluxetable}

%%%%%%%%%%%%%%%%%%%%%%%%%%%%%%%%%%%%%%%%%%%%%%%%%%%%%%%%%%%%%%%%%%%%%%

%% 1 %%
%%%%%%%%%%%%%%%%%%%%%%%%%%%%%%%%%%%%%%%%%%%%%%%%%%%%%%%%%%%%%%%%%%%%%%
%%%%%%%%%%%%%%%%%%%%%%%%%%%%%%%%%%%%%%%%%%%%%%%%%%%%%%%%%%%%%%%%%%%%%%
%%

%% 2 %%
%%%%%%%%%%%%%%%%%%%%%%%%%%%%%%%%%%%%%%%%%%%%%%%%%%%%%%%%%%%%%%%%%%%%%
% Western nucleus %%%%%%%%%%%%%%%%%%%%%%%%%%%%%%%%%%%%%%%%%%%%%%%%%%%

%%%%%%%%%%%%%%%%%%%%%%%%%%%%%%%%%%%%%%%%%%%%%%%%%%%%%%%%%%%%%%%%%%%%%
%%
 
%% 3 %%
%%%%%%%%%%%%%%%%%%%%%%%%%%%%%%%%%%%%%%%%%%%%%%%%%%%%%%%%%%%%%%%%%%%%%
% Eastern nucleus %%%%%%%%%%%%%%%%%%%%%%%%%%%%%%%%%%%%%%%%%%%%%%%%%%%
%
%%%%%%%%%%%%%%%%%%%%%%%%%%%%%%%%%%%%%%%%%%%%%%%%%%%%%%%%%%%%%%%%%%%%%
%%

%% 4 %%
%%%%%%%%%%%%%%%%%%%%%%%%%%%%%%%%%%%%%%%%%%%%%%%%%%%%%%%%%%%%%%%%%%%%%%
%%%%%%%%%%%%%%%%%%%%%%%%%%%%%%%%%%%%%%%%%%%%%%%%%%%%%%%%%%%%%%%%%%%%%%%
%%

%% 5 %%
%%%%%%%%%%%%%%%%%%%%%%%%%%%%%%%%%%%%%%%%%%%%%%%%%%%%%%%%%%%%%%%%%%%%%%
%%%%%%%%%%%%%%%%%%%%%%%%%%%%%%%%%%%%%%%%%%%%%%%%%%%%%%%%%%%%%%%%%%%%%%
%%

%% 6 %%
%%%%%%%%%%%%%%%%%%%%%%%%%%%%%%%%%%%%%%%%%%%%%%%%%%%%%%%%%%%%%%%%%%%%%%
%%%%%%%%%%%%%%%%%%%%%%%%%%%%%%%%%%%%%%%%%%%%%%%%%%%%%%%%%%%%%%%%%%%%%%
%%

%% 7 %%
%%%%%%%%%%%%%%%%%%%%%%%%%%%%%%%%%%%%%%%%%%%%%%%%%%%%%%%%%%%%%%%%%%%%%%

%%%%%%%%%%%%%%%%%%%%%%%%%%%%%%%%%%%%%%%%%%%%%%%%%%%%%%%%%%%%%%%%%%%%%%
%%

%% 8 %%
%%%%%%%%%%%%%%%%%%%%%%%%%%%%%%%%%%%%%%%%%%%%%%%%%%%%%%%%%%%%%%%%%%%%%%
%%%%%%%%%%%%%%%%%%%%%%%%%%%%%%%%%%%%%%%%%%%%%%%%%%%%%%%%%%%%%%%%%%%%%%
%%

%% 9 %%
%%%%%%%%%%%%%%%%%%%%%%%%%%%%%%%%%%%%%%%%%%%%%%%%%%%%%%%%%%%%%%%%%%%%%%
%%%%%%%%%%%%%%%%%%%%%%%%%%%%%%%%%%%%%%%%%%%%%%%%%%%%%%%%%%%%%%%%%%%%%%
%%

\end{document}